\documentclass[manuscript,screen,nonacm]{acmart}
\usepackage{multirow}
\usepackage{algorithm}
  \usepackage{algpseudocode}
  \usepackage{graphicx}
\usepackage{subcaption}
\AtBeginDocument{%
  }

\newcommand{\priorwork}{FASM}
\newcommand{\dhat}{\hat{d}}
\begin{document}


% \title{A Multi-Objective Optimization Framework for Variation-Aware Buffering of Clock Meshes}
\title{A Surrogate Model for Variation-Aware Clock-Mesh Analysis and Design-Space Exploration using Graph Neural Networks}
\footnotetext[1]{A preliminary version of this article was accepted at the 2026 ACM Great Lakes Symposium on VLSI (GLSVLSI'26) which we refer to as \priorwork{}~\cite{khan:gatmesh}.}

\author{Muhammad Hadir Khan}
\email{mkhan33@ucsc.edu}
\orcid{0000-0002-8136-9410}
\affiliation{%
  \institution{University of California, Santa Cruz}
  \city{Santa Cruz}
  \country{USA}
}

\author{Wajid Ali}
\email{wali2@ucsc.edu}
\affiliation{%
  \institution{University of California, Santa Cruz}
  \city{Santa Cruz}
  \country{USA}
}

\author{Arya Thottiparambil Manikandan}
\affiliation{
\institution{University of California, Santa Cruz}
\city{Santa Cruz}
\country{USA}
}

\author{Matthew Guthaus}
\affiliation{%
  \institution{University of California, Santa Cruz}
  \city{Santa Cruz}
  \country{USA}
}

\renewcommand{\shortauthors}{Muhammad H. Khan et al.}

\begin{abstract}
 Clock meshes are widely used in high-performance VLSI systems to minimize skew and tolerate On-Chip Variation (OCV) in nanometer technologies. However, analyzing meshes is difficult because reconvergent paths and multi-source drivers prevent static timing analysis (STA) tools from handling mesh loops, making SPICE the industrial standard. SPICE simulations are accurate but slow, so clock-mesh synthesis relies on approximate models that overlook critical effects such as
non-linear buffer driving caused by input slew and input skew. We propose a Graph Neural Network (GNN) based surrogate model of the clock mesh with a new graph representation and a new architecture that predicts power while also improving the delay and slew predictions at the sinks and capturing the skew histogram under OCV. Trained on SPICE data and tested on unseen real designs, our model achieves average errors of $3.40$~ps in delay, $4.14$~ps in slew and $4.46$~ps in skew, compared with $6.28$~ps, $12.35$~ps, and $11.62$~ps for our prior model, and $74.37$~ps delay and $27.51$~ps skew for the approximate method used in current state-of-the-art works. Our model predicts power within $3.4\%$ of SPICE on average, while delivering $18\times$-$1706\times$ speed-ups over the approximate method under OCV modeling, enabling faster and more accurate clock-mesh design-space exploration. A case study on a real design further shows that our surrogate model finds buffering candidates that, at a balanced tradeoff, improves $99^{\text{th}}$-percentile skew by $49.1\%$, reduces power by $25.7\%$, and reduces area by $37.3\%$ at
the cost of a $6.8\%$ increase in $99^{\text{th}}$-percentile maximum slew as compared to the state-of-the-art work.
\end{abstract}


\maketitle
\section{Introduction}\label{sect:intro}
\begin{figure}[t]
    \centering
    \includegraphics[width=0.8\columnwidth, height=4cm,keepaspectratio]{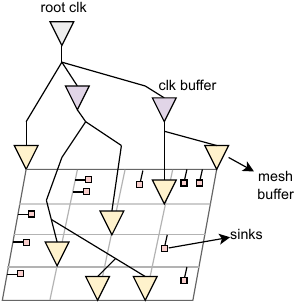}
    \caption{A clock tree driving the root clock to the mesh buffers through clock tree buffers. The redundant paths improve the tolerance to variations.
    \label{fig:mesh}}
\end{figure}

Clock Distribution Networks (CDNs) are fundamental to high-performance VLSI
designs, ensuring precise synchronization across millions of sequential
elements. Among various architectures, clock meshes stand out for their
superior robustness to On-Chip Variation (OCV),
offering low clock skew and improved tolerance to uncertainties in modern
fabrication processes~\cite{power4_clock, power6_clock, desai:meshsizing, amd:piledriver, amd:bulldozer}. By leveraging multiple redundant paths, as shown in Fig.~\ref{fig:mesh}, clock meshes
robustly distribute the clock signal across the entire chip. However, these
advantages come with significant challenges.

The difficulty in analyzing clock meshes arises from the
simultaneous multi-source driving of the network. These sources originate from
an underlying clock tree, where variations in delay among different branches
introduce varying input arrival times to the mesh. Such input skew,
combined with the inherently reconvergent paths within the mesh, means no independent driver-to-sink timing
arcs exist, so traditional static timing analysis (STA) cannot derive mesh sink timing. SPICE-based simulation remains the industrial fallback, but is computationally prohibitive at scale and for repetitive calling inside mesh synthesis or design-space exploration.
% \todo{In mesh, show buffered tree driving grid buffers driving a mesh with the sinks.}

The primary objective decides the requirements of a clock mesh analysis tool. For timing sign-off, accuracy is the only requirement, and for that SPICE remains unmatched. For clock mesh synthesis and design-space exploration, SPICE is infeasible since the evaluator must be called thousands of times on different candidates. So, four requirements follow. \textbf{Req. 1:} \textit{the evaluator must model simultaneous multi-source driving with reconvergent paths natively}, since the mesh reconvergent structure provides its robustness to OCV and loop breaking is not trivial. \textbf{Req. 2:} \textit{the evaluator must be reusable across topology changes}, so that any per-mesh setup cost is not repaid at every candidate evaluation. \textbf{Req.~3:}~\textit{it must be able to represent the input skew and slew perturbations}, since the OCV on the top-level tree that drives the mesh buffers is a dominant source of mesh skew. \textbf{Req. 4:} \textit{it must annotate delay, slew and power}, since skew (derived from delays) and power are optimization objectives and slew is a constraint. 

Prior methods satisfy partial requirements. The dedicated clock mesh analyzers~\cite{mesh:window-scheme, mesh:harmonic-fft, mesh:stat-mesh} model multi-source driving correctly and are considerably more accurate than closed-form approximations satisfying Req. 1, but each amortizes a large per-mesh setup that buffering changes invalidate, therefore not meeting Req. 2. The learning-based interconnect timing models~\cite{gnn-wire-timing, Cheng2020:WireTiming, gnn-synctree, gnn-timing-engine} are fast and reusable but assume a net driven by a single cell and a directed acyclic structure, thus not meeting Req. 1. The mesh loops where handled, are broken by node splitting~\cite{Cheng2020:WireTiming}, discarding the reconvergence that gives a mesh its variation tolerance. Consequently, current clock mesh synthesis frameworks~\cite{meshworks, clockmesh:dp, cmesh-int} annotate delays with the first-moment (FM) of the impulse response and estimate power with switched capacitance model $CV^2f$. FM satisfies Req. 1 and Req. 2 by representing the entire mesh structure and requiring only a single sparse matrix solve. However, it does not meet Req. 3 and Req. 4 by implicitly assuming all buffers switch simultaneously, so OCV-induced input skew cannot be represented and the method yields no transition time (slew) at sinks. Thus, prior works either offer accuracy without design-space exploration usability, or usability without variation awareness and power modeling.  

% Related work on clock meshes and GNNs
Our prior work~\cite{khan:gatmesh} proposed the first GNN based model for clock mesh timing analysis that annotates delay and slew at sinks in a fast manner. It satisfies Req.~1, Req.~2, and Req.~3 but not Req.~4 since it does not perform power analysis. Furthermore, OCV analysis is limited with only 500 MC simulations on two designs and missing the skew histogram tail metrics and plots. Moving forward we refer to our prior work as \priorwork~\cite{khan:gatmesh} and propose an extension to this work by creating a new graph representation and model architecture to support power prediction and significantly improved delay/slew accuracy, followed by a methodology to capture the tail-ends of the skew histogram under OCVs, and a synthetic benchmark generation process that improves model generalization in order to be used inside the design-space exploration. Our contributions are to:  

\begin{itemize}
    \item Propose a new graph representation that better encapsulates the clock mesh structure and improves performance.
    \item Propose a new model architecture for better delay, slew, skew, and power predictions under OCVs.
    \item Create a new synthetic benchmark generation strategy that improves model generalization for design-space exploration.
    \item Propose the Active Learning (AL) strategy along with a tail-weighted loss function that helps capture the tail-ends of the skew histogram under OCV. 
    \item Provide a case study of our proposed model used inside an optimization loop to provide tradeoffs for performance, power, and area metrics. 
  
\end{itemize}
The rest of the article is organized as follows: Section~\ref{sect:bg-related-work} provides background and brief overview of the related work. Section~\ref{sect:surrogate} explains our surrogate model for clock mesh analysis along with its training framework. Section~\ref{sect:tail-end-methodology} discusses the methodology for capturing the tail-ends of the skew histogram under OCVs followed by Section~\ref{sect:experiment-setup} and Section~\ref{sect:results} that explains our baseline comparison setup and results respectively. Section~\ref{sect:case-study} shows a case study using our surrogate model for buffering optimization and finally Section~\ref{sect:conclusion} presents the conclusion.

\section{Background and Related Work}\label{sect:bg-related-work}
\subsection{Clock Meshes}
Clock meshes are defined as a grid of $m\times n$ wires where $m$ represents the rows and $n$ represents the columns. At each of the $m\cdot n$ points where the horizontal and vertical wires intersect is a \textit{mesh node}.  The destinations of the clock signal are the \textit{sinks} which are the flip-flop clock pins. The sinks attach to the mesh wires through small wires called \textit{stubs}. Each mesh node may be driven by a buffer or not which depends on the sink density near the mesh nodes. We call these buffers the \textit{mesh buffers}. The clock skew of the mesh is defined as the difference between the maximum and minimum sink arrival times, $\text{skew} = \max_i d_i - \min_i d_i$, where $d_i$ is the delay at sink $i$. The sinks at the extremes of this delay distribution i.e, the fastest and slowest sinks within a mesh, are referred to as the \textit{tail sinks}, since they reside in the tails of the per-mesh delay distribution.

\subsection{STA Limitations and the Cost of SPICE}
% brief overview of how sta for delay annotations work along a path and why we cannot use a similar approach for clock-meshes due to reconvgering paths.
In traditional STA tools like Synopsys PrimeTime, a given netlist is converted to a \textit{timing graph} which is a directed-acyclic-graph (DAG). The delays are propagated based on precharacterized lookup-tables for each library gate assuming a single driver charging an effective capacitive load. Clock meshes do not follow this assumption where each capacitive load is driven by many drivers simultaneously. Thereby, no independent driver-to-sink timing arcs exists and STA cannot derive the clock mesh sink timing. Hence, it relies on per-sink delay and slew annotations obtained from full SPICE simulation which is accurate but inefficient due to runtime. Also, its cost compounds under variation. To verify clock mesh timing, thousands of Monte-Carlo (MC) simulations are run by varying die-to-die and within-die parameters to check whether the mesh satisfies skew constraints under OCV~\cite{clock-mesh-spice}. If the mesh does not, multiple iterations of optimization follow, each requiring the same verification, thereby consuming huge design turnaround time.

We now assess prior work against the requirements established in Section~\ref{sect:intro}.
\subsection{Dedicated Clock Mesh Analyzers}
The methods described here are built specifically to analyze clock meshes and so they satisfy Req. 1 by construction. Chen et al.~\cite{mesh:window-scheme} slide a window across the mesh and run SPICE on the netlist captured in each window, reporting agreement within $1\%$ of full mesh SPICE on clock latency. Ye et al.~\cite{mesh:harmonic-fft} apply harmonic-weighted model order reduction with port sliding, motivated by the steady-state operation of the clock networks. Standard reduction fails here because the mesh drivers introduce a larger number of ports, and those ports are the driver output themselves. Hao and Shi~\cite{mesh:stat-mesh} apply Kron's branch tearing recursively, encoding every resistive link and every non-root source resistor as a level of a link-BDD evaluated against a tree-BDD, at $O(N(L+S)^2)$ cost in nodes $N$, links $L$, and sources $S$.

None of the three methods meet Req. 2. Chen et al. regenerate SPICE per analysis, reporting hours for a single $65\times65$ mesh. Ye et al.~\cite{mesh:harmonic-fft} take the driver outputs as the ports of their reduced model, so any inserted, removed, or relocated buffer forces reconstruction. Hao and Shi~\cite{mesh:stat-mesh} make the source count a structural parameter of the decomposition, and their BDD construction dominates runtime, growing $639\times$ from a $5\times5$ mesh to a $15\times15$ mesh.

\subsection{Learning-Based Interconnect and Clock Timing}
This line of work applies learning to timing estimation and satisfies Req. 2 naturally, since inference carries no per-instance setup. SyncTREE~\cite{gnn-synctree} predicts clock-tree timing accurately using GNNs through two-step message passing from sinks to drivers and back. Guo et al.~\cite{gnn-timing-engine} build a GNN around a clock-tree timing engine. Both the works show good accuracy but assume a directed acyclic graph with a single path from driver to sink since their motivation is clock-tree timing analysis. Therefore, they do not meet Req. 1.

Cheng et al.~\cite{Cheng2020:WireTiming} and Ye et al.~\cite{gnn-wire-timing} target non-tree RC nets and are therefore closer to our setting, but both assume a net driven by one driver. Cheng et al. break the loops by splitting nodes in breadth-first search order from a single source such that the Elmore delay is preserved at the split points. Ye et al. avoid loop breaking but define each wire path as the shortest route from source to sink, with node features including Elmore downstream capacitance and stage delay computed along that path. Their aggregation of path RC into compact features is somewhat similar to our edge attributes (discussed in Section~\ref{sect:graph-representation}), and is motivated by the same concern with GNN depth and oversmoothing. The major difference is that a single source makes the path unique, so downstream capacitance and stage delay are well defined. Whereas, in a mesh every sink has a path from every buffer, and our edge attributes carry raw accumulated RC from each candidate driver rather than an Elmore quantity presupposing a direction of propagation. Additionally, both of these works obtain the driving cell's delay from library lookup tables.

Neither assumption transfers to a clock mesh, where every sink is driven simultaneously by many buffers, so downstream capacitance is undefined and the buffer delay itself is a difficult quantity to measure rather than a simple table lookup, thus not satisfying Req. 1. Tearing the loops also destroys redundancy that gives mesh its variation tolerance. On that note, Ye et al.~\cite{gnn-wire-timing} report that loop breaking proposed by Cheng et al.~\cite{Cheng2020:WireTiming} degraded non-tree accuracy from an $R^2$ of $0.99$ to $0.67$. Since these models admit a single source arrival time, they are unable to represent input skew across multiple drivers and hence they do not satisfy Req. 3 either. 

\begin{figure}[t]
    \centering
    \includegraphics[width=0.8\linewidth]{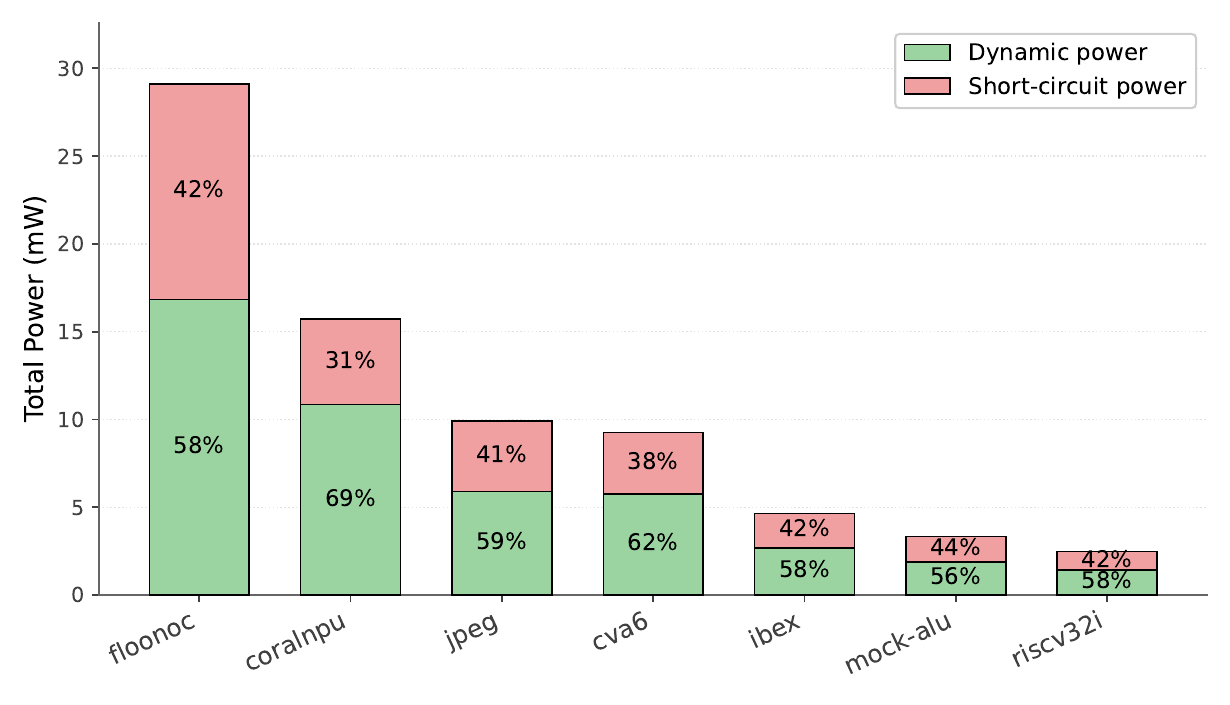}
    \caption{Breakdown of total power consumption in clock meshes. The short-circuit power contributes significantly to the total power which is obtained by running SPICE.}
    \label{fig:short-circuit-pow}
    % \vspace{-0.4cm}
\end{figure}

\subsection{Approximations used in Mesh Synthesis}\label{sect: related-work-fm}
Since no accurate clock mesh evaluators satisfy Req. 2, current mesh synthesis and optimization frameworks~\cite{meshworks, clockmesh:dp, cmesh-int} annotate delays using the first-moment (FM) of the impulse response and estimate power with a switched capacitance $CV^2f$ model. FM satisfies Req. 1 and Req. 2 where it models the mesh as a whole, and each evaluation is a single sparse matrix solve, making it the fastest method in the literature. However, it has some limitations:

\subsubsection{Delay annotations through FM method} The FM method computes delay from the first moment of the impulse response, which is obtained by forming a DC equivalent circuit of the mesh. This form excites all buffers from a common source implicitly assuming simultaneous switching. However, OCV induces different arrival times per buffer, thereby limiting its representational capability. Also, it cannot compute the output transition time and hence sink slews are unavailable. We cover the details of the FM method in Section~\ref{sect:experiment-setup} and show its performance in Section~\ref{sect:results}.

\subsubsection{Power optimization through dynamic power estimate} Dynamic power alone misses a key additional power component which is the short-circuit power. As noted by Wilke et al.~\cite{wilke:buffer} it has a significant portion of the total power consumption in clock meshes. This is caused by the multi-source structure of the mesh and happens due to OCV induced input skew at the mesh buffers, different output slews because of buffer size or load mismatch, or due to different drive strengths contending on the same mesh segment. A short-circuit current can flow directly between the buffers through the mesh wires in a brief window during which one driver might be pulling logical high while the other is still pulling logical low. To motivate the power analysis, Fig.~\ref{fig:short-circuit-pow} shows the contribution of short-circuit power in clock meshes of real designs from the HighTide benchmarks~\cite{hightide-bench} operating on $1$~GHz frequency and $0.7$ nominal VDD. The total power is computed through SPICE by taking the integral of current over a full clock period. Short-circuit power is obtained as the difference between total measured power and the dynamic power estimate. It constitutes a significant portion of the total power so a $CV^2f$ estimate underestimates the total power.

State-of-the-art works annotate delays with FM even though more accurate evaluators exist, precisely because those evaluators cannot be invoked for per-candidate evaluation during the search phase. It is also the only method that represents simultaneous multi-driving structure of the mesh while retaining fast evaluation. \textit{This is the reason why we adopt FM as our primary baseline in this work.}

The above discussion shows that no prior work satisfies all four requirements. \textit{This gap motivates our work to propose a surrogate that inherently captures the clock mesh structure, supports the OCV induced input skew and slew perturbations, and still provide accurate delay, slew, and power annotations in a faster way to be invoked inside STA timing closure and inside an optimization loop for design-space exploration.}

\subsection{Graph Neural Networks}
GNNs have seen tremendous adaptability in EDA where learning on an unstructured netlist is essential. In general, GNNs follow a message passing framework where each node $u \in V$, where $V$ is a set of nodes, aggregates information from its neighborhood $N(u)$:
\begin{equation} \label{eq:message-passing}
h_u^{(k+1)} = U^{(k)}(h_u^{(k)},AGG^{(k)}(\{h_v^{(k)}, \forall v \in N(u)\}))
\end{equation}
where $AGG(\cdot)$ generates a message from the aggregated neighborhood of $u$ and $U(\cdot)$ combines it with the current embedding $h_u^{(k)}$. Each such iteration $k$ is a layer of the GNN, with the initial embeddings at $k=0$ set to the input node features.

Our work uses the Graph Attention Network (GAT)~\cite{gat}, which instantiates $AGG(\cdot)$ as an attention-weighted sum of the neighbors, $h_u^{k+1} = \sigma(\sum_{v \in N(u)} \alpha_{u,v} W h_v)$, with attention weights
\begin{equation} \label{eq:gat-attn}
    \alpha_{u,v}
= \frac{
    \exp\Bigl(\sigma \bigl(a^\top \bigl[W h_{u} \oplus W h_{v}\bigr]\bigr)\Bigr)
  }{
    \sum_{v' \in N(u)} \exp\Bigl(\sigma\bigl(a^\top \bigl[W h_{u} \oplus W h_{v'}\bigr]\bigr)\Bigr)
  }
\end{equation}
where $\sigma$ is a non-linear activation, $a$ is a trainable attention vector, $W$ is a trainable weight matrix, and $\oplus$ denotes concatenation. This assigns each neighbor an importance in $[0,1]$. As in the transformer architecture~\cite{transformer}, GAT uses $H$ independently parameterized attention heads whose outputs are averaged. For graph-level regression, the node embeddings must additionally be summarized into a single vector through pooling, either global~\cite{globalpool} or hierarchical~\cite{hierpool}, before a Multi-Layer Perceptron (MLP) produces the prediction.

\section{Surrogate Model for Clock Mesh Analysis}\label{sect:surrogate}
\subsection{Graph Representation}\label{sect:graph-representation}
\begin{figure}[t]
    \centering
    \includegraphics[height=3cm, keepaspectratio]{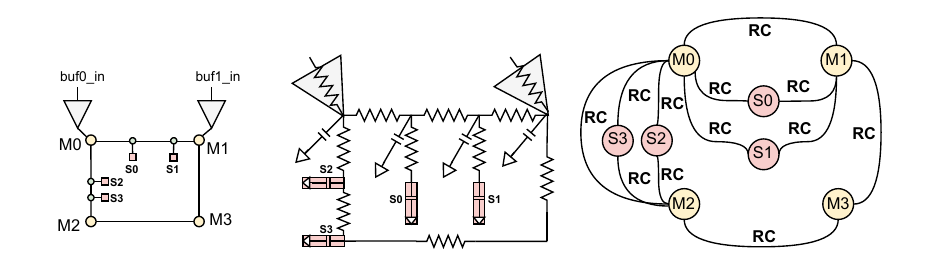}
    % \vspace{-0.3cm}
    \caption{\textit{Left:} A trivial clock-mesh grid example with buffering and sinks. \textit{Middle:} Equivalent RC mesh network. \textit{Right:} Our graph model representation.}
    \label{fig:graph-model}
\end{figure}

In this work, we represent the clock mesh as a bidirected homogeneous graph $G=(V, E)$ with edge attributes as shown in Fig.~\ref{fig:graph-model}. The mesh wires are represented as a standard $\pi$-model with lumped capacitance and resistance which are used as edge features in the graph. The \textit{sinks} are modeled as load capacitance to ground. The \textit{mesh nodes} can either be buffered or not. In the case of Fig.~\ref{fig:graph-model}, \textit{mesh nodes} M0 and M1 are buffered whereas M2 and M3 are not. As compared to \priorwork{} this representation helps us in the following way:

\textbf{Wire RC as edge features eliminates the wire nodes:} FASM’s representation relies on node features alone where every mesh wire segment becomes a wire node carrying its resistance and capacitance. Encoding that RC as an edge attribute instead removes those nodes entirely, thereby shrinking our graphs. 

\textbf{Aggregating the mesh-to-sink path RC into one edge eliminates auxiliary connections:} Even with RC on edges, a sink separated from its potential driver by several wire segments still requires several rounds of message passing, requiring deep GNNs that degrade performance due to oversmoothing~\cite{oversmoothingsurvey}. Previous works have shown GNNs in EDA have seldom stacked beyond 4 layers without oversmoothing and overfitting~\cite{gnn-timing-engine}. \priorwork{} worked around this by adding auxiliary nodes to provide direct mesh-to-sink path. We instead accumulate the total path RC from a mesh node to each sink it may drive as an edge attribute, so a sink is one hop away from its potential driver and the auxiliary connections are unnecessary. 

Using this representation removes the wire nodes and the auxiliary nodes thereby significantly reducing the total number of nodes in the graph, improving timing metric predictions, and providing faster inference time later discussed in Section~\ref{sect:results}.

% NEW TABLE
\begin{table}[t]
\centering
\caption{Node/Edge features and labels.}
\setlength\tabcolsep{3pt}
\vspace{-0.3cm}
\begin{tabular}{lrr}
\toprule
\textbf{Type} & \textbf{Name (units)}  & \textbf{Description}  \\
\midrule
\multirow{8}{*}{Node features} & input\_delay (ps)     & Buffer input delay from tree \\
& input\_slew (ps)   & Buffer input slew from tree  \\
& buf\_cap (fF)    & Buffer effective capacitance  \\
& buf\_res (ohms)      & Buffer effective resistance  \\
& buf\_len (nm) & Buffer channel length \\
& buf\_vdd (V) & Buffer VDD supply \\
& buf\_vth (V) & Buffer threshold voltage \\
& sink\_cap (fF)  & sink load capacitance  \\
& x (nm) & normalized x-coordinate \\
& y (nm) & normalized y-coordinate \\
& is\_sink & is this a sink node \\
& is\_mesh\_node  & is this a grid mesh node \\
& is\_buffered & is this mesh node buffered \\
\midrule 
\multirow{2}{*}{Edge features} & wire\_res  & wire resistance \\
& wire\_cap & wire capacitance \\
\midrule
\multirow{5}{*}{Global features} & total\_mesh\_cap & sum of wires, sink, buffer capacitance \\
& sink\_density & ratio of sinks to chip area \\
& buffer\_density & ratio of buffers to total mesh nodes \\
& max\_stub\_length & the max stub wire length in the grid \\
& min\_stub\_length & the min stub wire length in the grid \\
\midrule
\multirow{2}{*}{Node labels} & delay\_out (ps) & delay value at each sink \\
& slew\_out (ps) & slew value at each sink \\
\midrule
Graph label & power (mW) & total power of mesh \\
\bottomrule
\end{tabular}
\label{tab:node-features}
\end{table}

\subsection{GNN Features and Labels}\label{sect:gnn-feat-labels}
Each node in our graph has an associated feature vector as shown in Table~\ref{tab:node-features}. The features \textit{is\_mesh\_node} and \textit{is\_sink} are one-hot encoded features that help the GNN to identify \textit{mesh nodes} (M) and \textit{sinks} (S). Also, each \textit{mesh node} can be buffered or not, represented by the \textit{is\_buffered} feature. Note that in homogeneous GNNs all nodes are treated similarly, but we refer to nodes as M or S just for the reader's convenience. All nodes have a feature vector with fixed dimension $\vec{x} \in R^d$. Any feature that does not apply to a relevant node has default value of $0$. The \textit{input\_delay} and \textit{input\_slew} capture the delay and slew of the top-level tree arriving at the \textit{mesh buffers} (see Fig.~\ref{fig:mesh}). 
The \textit{buf\_cap} and \textit{buf\_res} capture the driving strength of the buffer and help learn its internal delay given the \textit{input\_slew} and the load it is driving through the \textit{sink\_cap} and the \textit{wire\_res, wire\_cap} features. 
\begin{figure}[t]
    \centering
    \includegraphics[height=2.5cm, keepaspectratio]{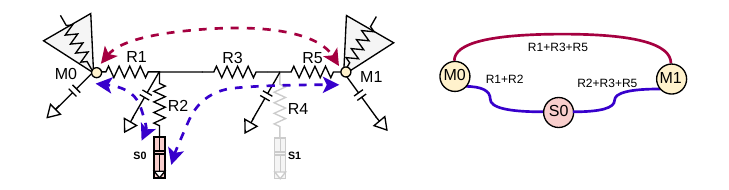}
    % \vspace{-0.3cm}
    \caption{Calculation of the RC edge features between sink (S0) and its two nearby mesh nodes (M0, M1) and between two adjacent mesh nodes. (S1) follows similar pattern. }
    \label{fig:edge-feat}
\end{figure}
We create edges between each \textit{sink} and its two nearest \textit{mesh nodes}. This edge has a feature of the path RC from the \textit{sink} to this \textit{mesh node} and vice-versa as shown in Fig.~\ref{fig:edge-feat}. For path S0 to M1 we add R2+R3+R5, stub resistances of other sinks (R4) are excluded because they are not on the path and are usually minimal. In a similar manner, we also accumulate the path capacitances from each \textit{mesh node} to \textit{sink} and vice-versa (not shown in Fig.~\ref{fig:edge-feat} for clarity). This helps the \textit{sink} to identify whether it is driven significantly by a nearby \textit{mesh node} and the driving strength of the buffer. It also provides information to the \textit{mesh node} of how many \textit{sinks} it must drive and their load capacitance. We also create edges between the adjacent \textit{mesh nodes} as shown in Fig.~\ref{fig:edge-feat} where M0 and M1 are connected. This edge  provides insight for the GNN model of an unloaded mesh wire without any \textit{sinks} and only includes the mesh wire RC. In addition to the node and edge features we also utilize global features in order to encode global representations of the mesh. These features help our downstream model network to learn important global characteristics that impact the clock skew and power of a given mesh.

\begin{figure}[t]
    \centering
    \includegraphics[width=1.0\textwidth]{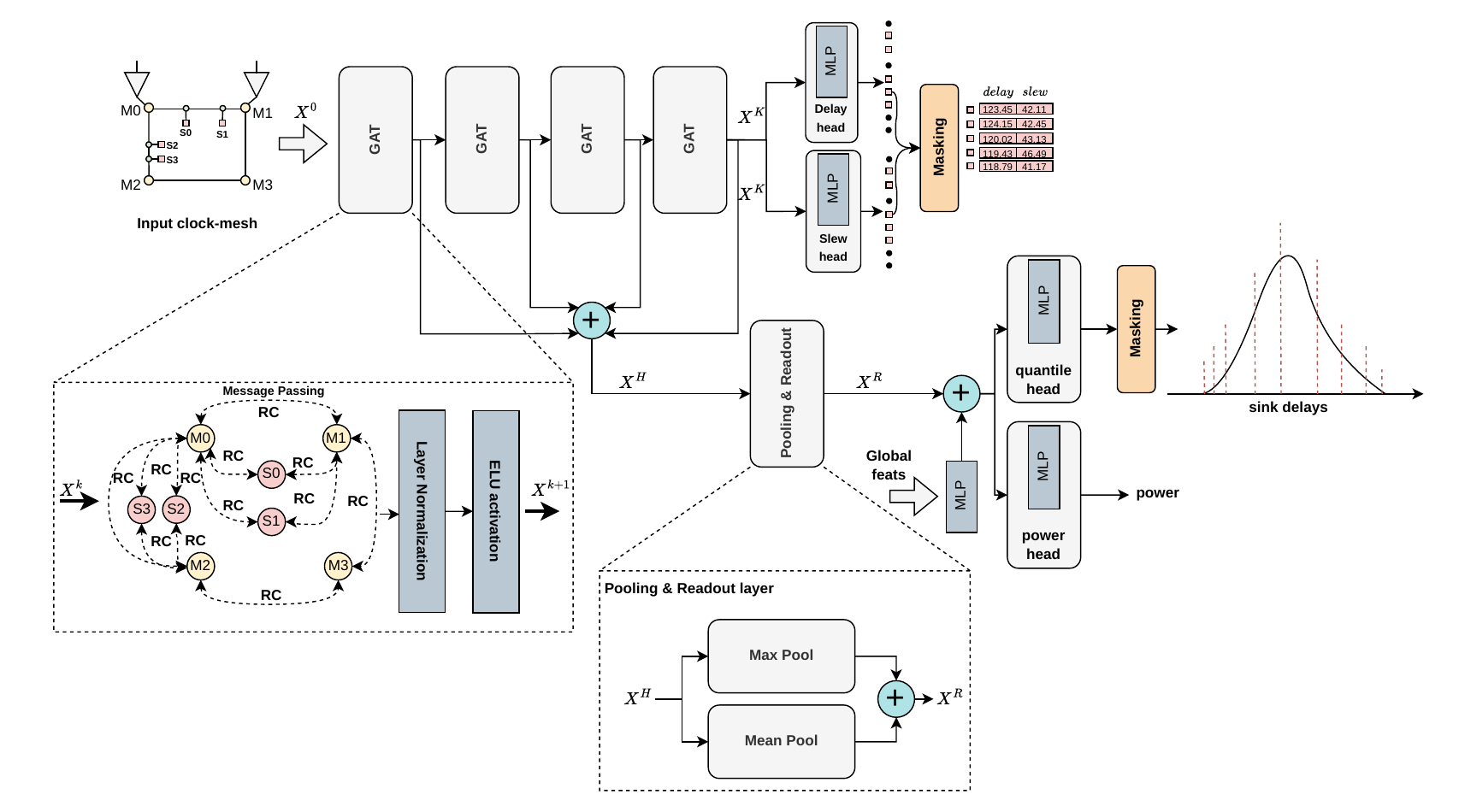}
    % \vspace{-0.3cm}
    \caption{Our model architecture uses $\boldsymbol{K}=4$ GAT layers to perform message passing followed by normalization and ELU activation. Final node embeddings $\boldsymbol{X^K}$ are used by separate MLP heads for delay and slew prediction and pooled to perform global tasks such as delay quantile predictions and power prediction. }
    \label{fig:model}
    % \vspace{-0.4cm}
\end{figure}

\subsection{Model Architecture}\label{sect:model-architecture}
We adopt a multitask learning framework~\cite{multitask-learning} which consists of delay and slew prediction tasks at each \textit{sink} as well as global tasks like delay distribution quantiles and power prediction. The backbone GNN is shared and we have task specific MLP heads as shown in Fig.~\ref{fig:model}. For global tasks we concatenate the shared encoder embeddings from each layer of the GNN ($X^H$) and pass it through the pooling and readout layer to get $X^R$. To provide additional global context we append the global features by projecting it through an MLP. This is so that the global embeddings are in a higher dimension and preserve the global context when appended with embeddings already in a higher dimension. These embeddings are then passed to the separate task specific heads (quantile and power). The quantile head predicts the quantiles of the sink delay distribution. This is useful as it helps the model to learn global embeddings that shape the delay distributions of the clock mesh sinks as well as the skew of the clock mesh. We specifically predict the $0\%, 1\%, 5\%, 10\%, 25\%, 50\%, 75\%, 90\%, 95\%,99\%,$ and $100\%$ quantiles. Although skew depends only on the $0^\text{th}$ and $100^\text{th}$ quantiles, supervising the intermediate quantiles regularizes shared readout where along with the extremes it should also learn to predict the bulk of the distribution. The power head predicts the power of the clock mesh by utilizing the same shared embeddings. The usefulness of this multitask learning architecture is that the shared GNN encoder learns node embeddings in a global context by utilizing the mesh topology and spatial patterns whereas task specific heads focus on objective specific learning. Since the delay, slew and power are all physically related by the RC network and its transient response, this acts as an inductive bias and regularizes all tasks thereby improving generalization of our model. 

We utilize GAT convolutional layers~\cite{gat} for our GNN implementation. However, the vanilla GAT convolution layers do not support edge features and rely on node features. For this, we modified~\eqref{eq:gat-attn} and compute attention scores by involving edge features as:
\begin{equation} \label{eq:our-attn}
    \alpha_{u,v} 
= \frac{ 
    \exp\Bigl(\sigma \bigl(a^\top \bigl[W h_{u} \oplus W h_{v} \oplus W_e\ e_{u,v}\bigr]\bigr)\Bigr)
  }{
    \sum_{v' \in N(u)} \exp\Bigl(\sigma\bigl(a^\top \bigl[W h_{u} \oplus W h_{v'} \oplus W_e \ e_{uv'}\bigr]\bigr)\Bigr)
  }
\end{equation}
% HYPERPARAMETER TABLE
\begin{table}[t]
\centering
\caption{Model Parameters}
\label{table:hyperparameters}
\vspace{-0.3cm}
\begin{tabular}{l|r|r|r}
\toprule
\textbf{Parameter} & \textbf{Layers} & \textbf{Channels} & \textbf{Attn. Heads}  \\
\midrule
GAT         & 4 & 128 & 8 \\
Delay/Slew/Quantile/Power MLP  & 3 & 64 & N/A \\
Global Proj. MLP & 2 & 64 & N/A \\
\bottomrule
\end{tabular}
\end{table}
This allows each node to give importance to its neighbors based on both their node and edge features. Each GAT message passing layer is followed by Layer Normalization~\cite{layer-norm} which helps stabilize training by normalizing hidden node embeddings. For non-linear activations we use ELU~\cite{elu}. A masking layer is used to omit nodes that are not sinks to calculate loss contributed by only the sink nodes.

\subsection{Model Parameters}
The model architecture parameters are shown in Table~\ref{table:hyperparameters}. Based on empirical evaluations, we did not find an improvement in performance with more than $4$ GAT layers and $8$ attention heads. The delay, slew, quantile and power heads all use similar MLP parameters. The global projection MLP uses $2$ layers to project the global features to a higher dimensional hidden representation. All these parameters are based on the best performance of the model.

\subsection{Model Loss Functions}\label{sect:model-training} 
Our loss function has a total of four components: losses from delay head, slew head, quantile head, and power head. We now discuss these losses.

Let a mini-batch contain $G$ mesh graphs, and let $\mathcal{S}_g$ denote the set of clock sinks of graph $g$, so that $|\mathcal{S}_g|$ is its sink count. For sink $i$, let $d_i$ be the SPICE-measured delay and $\dhat_i$ the delay head prediction. The delay loss is the mean over graphs of the per-graph mean squared error,
   \begin{equation}
  \mathcal{L}_{\text{delay}}
    \;=\
    \frac{1}{G}\sum_{g=1}^{G}
      \frac{1}{|\mathcal{S}_g|}
      \sum_{i \in \mathcal{S}_g} \bigl(\dhat_i - d_i\bigr)^{2}
    \label{eq:delay-loss}
  \end{equation} 
The slew loss is also computed as a standard per-graph MSE.  
   \begin{equation}
  \mathcal{L}_{\text{slew}}
    \;=\;
    \frac{1}{G}\sum_{g=1}^{G}
      \frac{1}{|\mathcal{S}_g|}
      \sum_{i \in \mathcal{S}_g} \bigl(\hat{s}_i - s_i\bigr)^{2},
    \label{eq:slew-loss}
  \end{equation}
 where $s_i$ denotes the true SPICE-measured transition time at sink $i$ and $\hat{s}_i$ for the slew head prediction.

Our model also has a global regression quantile head which predicts, for each design $g$, the delay distribution at a fixed set of $D=11$ probability levels $\boldsymbol\tau = (0, 0.01, 0.05, 0.1, 0.25, 0.5, 0.75, 0.9, 0.95, 0.99, 1)$, chosen to sample the tails densely. We also train this head using the MSE loss, where $Q_g(\tau_d)$ denotes the empirical $\tau_d$-quantile of the true sink delays and $\hat q_{g,d}$ for the prediction.
  \begin{equation}
  \mathcal{L}_{\text{quant}}
    \;=\;
    \frac{1}{GD}\sum_{g=1}^{G}\sum_{d=1}^{D}
      \bigl(\hat{q}_{g,d} - Q_g(\tau_d)\bigr)^{2}.
    \label{eq:quant-loss}
  \end{equation}
Because $\tau_1=0$ and $\tau_D=1$ are the per-graph minimum and maximum sink delay, the predicted clock skew is read directly off the quantile head as:
  \begin{equation}
  \widehat{\text{skew}}_g \;=\; \hat{q}_{g,D} - \hat{q}_{g,1}.
    \label{eq:skew-from-quantiles}
  \end{equation}

The power head predicts total active power $\hat P_g$ per design and is likewise trained by mean squared error against the SPICE-measured active power of graph $P_g$, 
 \begin{equation}
     \mathcal{L}_{\text{power}} = \frac{1}{G}\sum_g(\hat P_g - P_g)^2
 \end{equation}
  
  The total training loss combines all the terms:
  \begin{equation}
      \mathcal{L} = \lambda_{delay}\cdot\mathcal{L}_{\text{delay}}
                  +\lambda_{slew}\cdot\mathcal{L}_{\text{slew}} + \lambda_{quant}\cdot\mathcal{L}_{quant} + \lambda_{power}\cdot\mathcal{L}_{power}
      \label{eq:total_loss}
  \end{equation}
Where, $\lambda_{delay},\lambda_{slew}, \lambda_{quant}, \text{and} \ \lambda_{power}$ are the weights of each loss. With experiments we found out the best weights for each loss to be $\lambda_{delay}=1.0, \ \lambda_{slew}=0.1, \ \lambda_{quant}=0.1, \ \text{and} \ \lambda_{power}=1.0$.

\subsection{Synthetic Data Generation}\label{sect:synth-data-gen}
The synthetic benchmarks in \priorwork{}~\cite{khan:gatmesh} are all buffered using set-covering heuristics from prior works~\cite{pengli:combinatorial, meshworks}. This trains the model on a narrow buffering distribution which is not ideal for design-space exploration where extremely under-buffered or over-buffered meshes can be explored during the search phase. Therefore, in this work we generate synthetic benchmarks by sampling buffering density from uniform distribution range $[0.1, 1.0]$ and picking buffer sizes with uniform sampling from the given buffer library. This allows a model to learn from a wider distribution of buffering and its effects on the clock mesh.

\begin{figure}[t]
    \centering
    \includegraphics[width=0.8\linewidth]{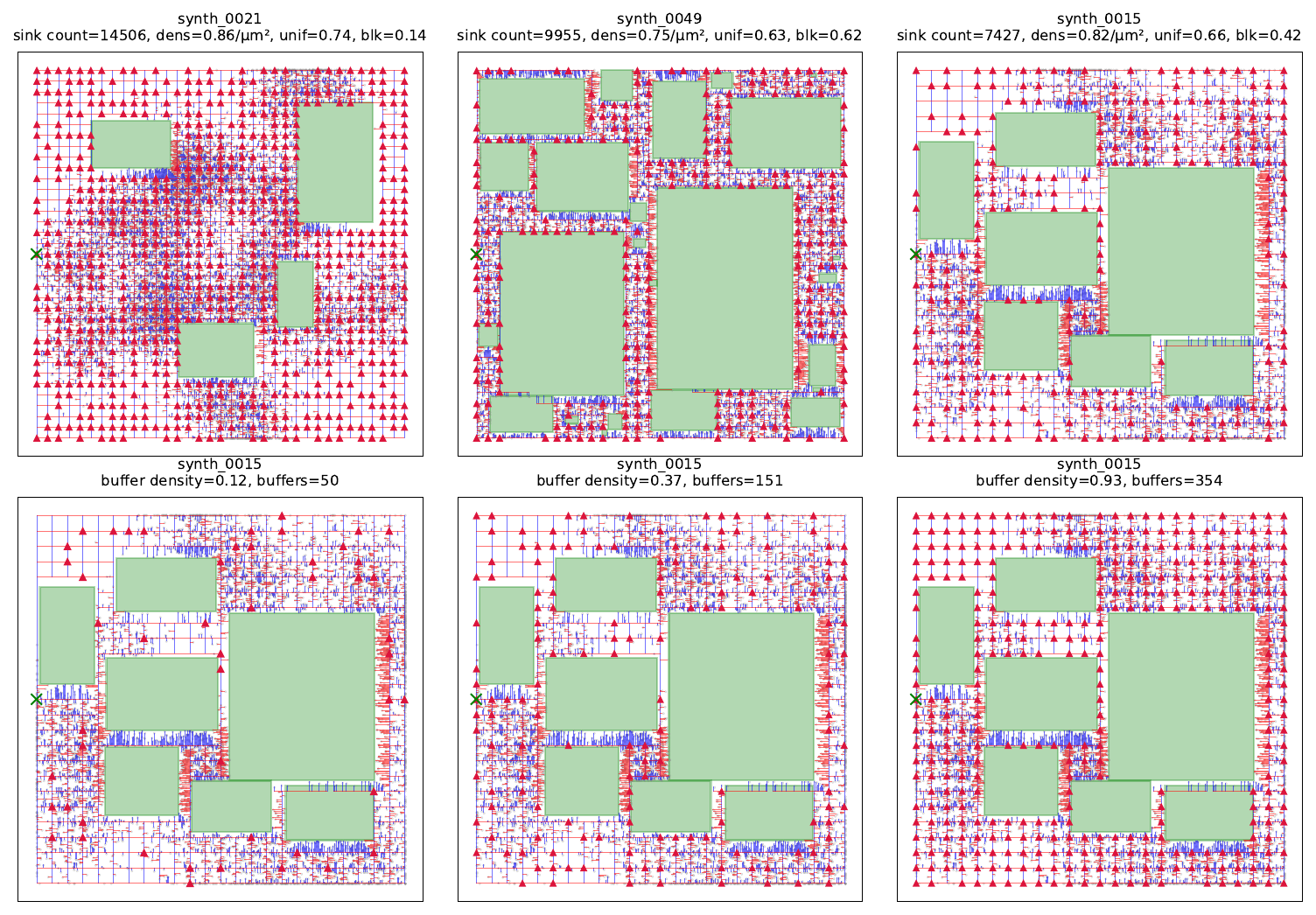}
    \caption{Synthetic Clock Meshes. Row 1: Shows different designs with varying statistics from Table~\ref{table:data-stats}. Row 2: Shows a single design with varying buffer densities. The green rectangles are the blockages and red triangles are buffers.}
    \label{fig:synth-meshes}
\end{figure}

In order to create synthetic benchmarks with realistic characteristics, we implemented open-source designs in the HighTide benchmarks~\cite{hightide-bench} using the OpenROAD~\cite{openroad} flow and extracted representative statistics of the clock meshes which are shown in Table~\ref{table:data-stats}. These four statistics jointly help describing the electrical behavior of the clock mesh. The number of sinks and the sink density describes the die area and how large the clock mesh would be created to drive the sinks. The spatial uniformity ratio characterizes how these sinks are distributed. It measures how the sinks are distributed across the die by comparing how far apart neighboring sinks are in real design to how far apart they would be under a random placement of the same sink count and die area. For each sink we take the distance to its nearest neighboring sink and average this over all sinks. We then divide by the mean nearest-neighbor distance expected if the same number of sinks were scattered uniformly at random over the die. This gives us a dimensionless value where if it is $< 1$ we have clustered sinks, with $\approx1$ one we have a random arrangement, and with $> 1$ we have regular spaced sinks. Lastly, the blockage ratio defines blockages where the clock mesh cannot be routed. Some of the generated synthetic benchmarks are shown in Fig.~\ref{fig:synth-meshes}.
\begin{table}[t]
\centering
\caption{Statistics for synthetic benchmark generation}
\label{table:data-stats}
% \setlength\tabcolsep{2pt}
% \vspace{-0.3cm}
\begin{tabular}{l|r|l}
\toprule
\textbf{Statistic} & \textbf{Ranges} & \textbf{Description}  \\
\midrule
sink count & $[500,15000]$ & Sets the size of the clock mesh. \\
% \midrule
sink density (sinks/$\mu m^2$) & $[0.037, 0.86]$ & Sinks per unit area; jointly with sink count it sets the die dimensions. \\
% \midrule
spatial uniformity ratio & $[0.4, 1.2]$ & Sink placement pattern, clustered ($<1$) random ($\approx 1$), or regular ($>1$).\\
blockage ratio & $[0.0, 0.65]$ & Fraction of the die area obstructed by blockages. \\

\bottomrule
\end{tabular}
% \vspace{-0.3cm}
\end{table}
 % SYNTHETIC CLOCK MESH IMAGES
% VARIATIONS TABLE
\begin{table}[t]
\centering
\caption{Variation Parameters for OCV. The (Affects) shows which variation parameter changes which graph node/edge feature.}
\setlength\tabcolsep{6pt}
\vspace{-0.3cm}
\begin{tabular}{lrr}
\toprule
\multicolumn{1}{l}{\textbf{Parameter (Affects)}} & \multicolumn{1}{r}{\textbf{Gaussian}$(\boldsymbol{\sigma} / \boldsymbol{\mu})$} & \multicolumn{1}{r}{\textbf{From}}  \\
\midrule
$L_{eff}$ (buf\_len) & $0.84/21$ (nm) & \textbf{\cite{Leff-Weff-var}} \\
$W_{eff}$ (buf\_cap/res,sink\_cap)& $0.26/6.5$ (nm) & \textbf{\cite{Leff-Weff-var}} \\
$V_{th}$ (buf\_vth) & $0.021/0.3$ (V) & \textbf{\cite{vth-var}} \\
$V_{dd}$ (buf\_vdd)  & $0.023/0.7$ (V) &  \textbf{\cite{asap7}} \\
Wire resistance (wire\_res) & $0.003/0.02$ ($\Omega/$nm) & \textbf{\cite{inteconnect-var}} \\
Wire capacitance (wire\_cap)& $5.4e^{-6}/1.8e^{-4}$ (fF$/$nm) & \textbf{\cite{inteconnect-var}} \\
\midrule 
\multicolumn{1}{l}{\textbf{Parameter (Affects)}} & \multicolumn{1}{r}{\textbf{Uniform}} & \multicolumn{1}{r}{\textbf{From}}  \\
\midrule
Clock tree Delay (input\_delay)  & $\pm50$ (ps) & \textbf{\cite{meshworks}}\\
Clock tree Slew (input\_slew)   & $\pm10$ (ps) & \textbf{\cite{meshworks}}\\
\bottomrule
\end{tabular}
\label{table:variation}
\end{table}
% ALGORITHM FOR SYNTHETIC BENCHMARK GENERATION
  \begin{algorithm}[t]
    \caption{Synthetic Benchmark Generation}
    \label{alg:synth-gen}
    \begin{algorithmic}[1]
      \Require Real designs $\mathcal{R}$; \#synthetic designs $N$; \#bufferings per design $P$; \#MC samples per buffering $M$; sigma multiplier $n_\sigma$; buffer library ${BL}$
    \Ensure Labeled clock-mesh dataset $\mathcal{D}$
    \Statex $n_{\text{sink}}$: number of clock sinks
    \Statex $\rho$: sink density, sinks per unit area
    \Statex $\gamma$: spatial uniformity ratio, $\gamma<1$ clustered, $\gamma>1$ regular
    \Statex $\beta$: fraction of die area covered by blockages
    \For{each axis $s \in \{n_{\text{sink}},\, \rho,\, \gamma,\, \beta\}$}
        \State $[\,s_{\min}, s_{\max}\,] \gets$ observed range of $s$ over $\mathcal{R}$
            \Comment{sampling ranges derived from real designs (Table~\ref{table:data-stats})}
    \EndFor

    \State $\Theta \gets \textsc{LatinHypercube}\big(N,\ \{[\,s_{\min}, s_{\max}\,]\}_s\big)$
        \Comment{uniform per-axis coverage}

    \State $\mathcal{B} \gets \emptyset$
    \For{each $(n_{\text{sink}}, \rho, \gamma, \beta) \in \Theta$}
        \State $A \gets n_{\text{sink}} / \rho$;\quad $\ell \gets \sqrt{A}$
            \Comment{die area and side length}
        \State $\mathcal{K} \gets \textsc{PlaceBlockages}(\beta, \ell)$
            \Comment{non-overlapping rects covering fraction $\beta$}
        \State $\mathcal{S} \gets \textsc{PlaceSinks}(n_{\text{sink}}, \gamma, \ell, \mathcal{K})$
            \Comment{match index $\gamma$, avoid $\mathcal{K}$}
        \State $\mathcal{B} \gets \mathcal{B} \cup \{\,(\ell, \mathcal{S}, \mathcal{K})\,\}$
    \EndFor

    \State $\mathcal{D} \gets \emptyset$
    \For{each benchmark $B \in \mathcal{B}$}
        \State $\{p_1,\dots,p_P\} \sim \mathcal{U}(0.1,\, 1.0)$
            \Comment{buffer densities drawn uniformly \emph{per design}}
        \For{$j \gets 1$ \textbf{to} $P$}
            \State $G \gets \textsc{SynthesizeMesh}(B,\; p_j,\; BL)$
                \Comment{mesh with buffering probability $p_j$, sizes from $BL$}
            \State $\mathbf{Z} \gets \Call{SampleVariation}{G,\; M,\; n_\sigma}$
            \ForAll{$\mathbf{z} \in \mathbf{Z}$} \textbf{in parallel}
                \State $y \gets \Call{RunSpice}{G,\; \mathbf{z}}$
                    \Comment{per-sink delay and slew, grid power}
                \State $\mathcal{D} \gets \mathcal{D} \cup \bigl\{\bigl(G,\; y\bigr)\bigr\}$
          \EndFor
        \EndFor
    \EndFor

    \State \Return $\mathcal{D}$
    \end{algorithmic}
    \end{algorithm}

Since each of the statistic can vary independently, this leads to an exponential number of synthetic designs needed to capture the entire space which quickly becomes infeasible due to the cost of clock-mesh synthesis and label generation by running SPICE simulations. Therefore, we use Latin Hypercube Sampling (LHS)~\cite{lhs} to generate our synthetic dataset. It treats the four statistics as independent axes bounded by the ranges observed in the real designs (shown in Table~\ref{table:data-stats}) and partitions each axis into equally probable intervals. Samples are drawn such that every interval on every axis is populated exactly once. This guarantees uniform coverage of each statistic individually while filling the joint space far more evenly than independent random sampling. 

Each synthetic benchmark is then synthesized into a clock mesh with a random buffering sweep to expose the model to the full range of buffering configurations. 

Finally, we run $M$ Monte-Carlo simulations by varying the parameters within their bounds listed in Table~\ref{table:variation} for each design to have variation induced labels with SPICE. This process is briefly described in the Algorithm~\ref{alg:synth-gen} where we set $N=50, P=20, M=5$ and $n_{\sigma}=3$ to apply $3\sigma$ variations. So in total we have $50\times20\times5=5000$ synthetic designs in the dataset. With $M=5$ we have only $5$ MC samples for each design to train the model on. Increasing $M$ significantly increases the training dataset whereas the effective modeling of $3\sigma$ tail-end skew histogram requires a large number of MC samples. We discuss this limitation and propose a solution in Section~\ref{sect:tail-end-methodology}.

\subsection{Training Methodology}\label{sect:train-methodology}
% IN TRAINING METHODOLOGY
To train the model, we do a train/validate/test split of 80/10/10 from our synthetic dataset generated by the process outlined in Section~\ref{sect:synth-data-gen}.
All the labels are generated using Synopsys HSPICE. For optimizing the model parameters, ADAM~\cite{adam-opt} was used with a learning rate of $7.5e^{-4}$ and weight decay of $5e^{-4}$ for L2-regularization. Training was done with batch size of $1$ and for a maximum of $5000$ epochs. Early-stopping was implemented to exit if the validation results did not improve for more than 10 epochs. The training of our model completed in $\approx9.4$ hours and took $56$ epochs before exiting early.

\begin{table}[t]
\centering
\caption{Rarity of tail-producing skews under uniform Monte Carlo
sampling. For each design, $5000$ variations were drawn and SPICE-simulated.
Where $f$ denotes the measured fraction of corners whose
skew reaches the mean$+3\sigma$ threshold. Since draws are independent,
$P(\text{at least one in } N) = 1-(1-f)^{N}$, giving $N=1/f$ on average and
$N=\ln(0.01)/\ln(1-f) \approx 4.6/f$ for $99\%$ confidence.}
\label{tab:tail_rarity}
\setlength{\tabcolsep}{5pt}
\begin{tabular}{lrrrrrr}
\toprule
 & skew mean & $\text{ skew (mean}+3\sigma)$ & \multirow{2}{*}{\# Samples $\geq3\sigma$} & \multirow{2}{*}{$f$} & \multicolumn{2}{c}{SPICE samples $N$} \\
\cmidrule(lr){6-7}
design & (ps) & (ps) & & & expected & $99\%$ conf. \\
\midrule
ibex     & 29.8 & 49.1 & 41 & 0.82\% & 122 &  559 \\
floonoc  & 33.3 & 46.7 & 34 & 0.68\% & 147 &  676 \\
jpeg     & 33.5 & 49.8 & 31 & 0.62\% & 161 &  740 \\
riscv32i & 31.0 & 54.9 & 28 & 0.56\% & 179 &  820 \\
mock-alu & 20.4 & 35.6 & 28 & 0.56\% & 179 &  820 \\
cva6     & 42.9 & 63.7 & 22 & 0.44\% & 227 & 1044 \\
coralnpu & 57.9 & 76.1 & 20 & 0.40\% & 250 & 1149 \\
\bottomrule
\end{tabular}
\end{table}

\section{Methodology for Tail-End OCV Improvement}\label{sect:tail-end-methodology}

The $3\sigma$ tail-end skews are important to analyze the robustness of clock mesh to OCVs. The mesh architecture, due to its reconverging and multiple buffer paths to each sink, intentionally smooths out most of the variation causing the tail-ends to occur in extremely rare variation patterns. In order to verify this, we ran $5000$ MC simulations and calculated the expected number of samples needed to get at least one sample that produces skew beyond mean+$3\sigma$ as shown in Table~\ref{tab:tail_rarity}. For example, if we want $99\%$ confidence of seeing at least one such sample then we need the number of samples listed under "$99\%$ conf." column in Table~\ref{tab:tail_rarity}. Thus, generating these tail events requires an extensive number of MC simulations and an exorbitant amount of time for SPICE label generation. Our model trained from synthetic data generation also misses this critical region due to being trained on only $5$ MC samples. We therefore further improve upon this model by proposing a new tail-aware loss function and an efficient Active Learning (AL) approach for targeted tail-end skew training. 

\subsection{Tail-Aware Loss Function}\label{sect:tail-loss}

\begin{figure}[t]
      \centering
      \includegraphics[width=0.5\columnwidth]{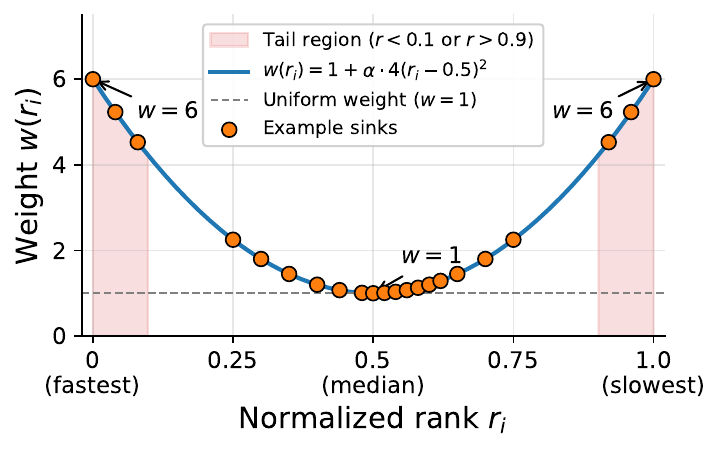}
      \caption{Proposed Tail-Aware Loss function. Each sink node is assigned a normalized rank $r_i \in [0, 1]$ based on its ground-truth delay. The function assigns higher weight to the fastest and slowest sinks (shaded tail regions) compared to median sinks. Orange dots represent example sink nodes in a mesh, where the majority cluster near the median with few sinks at the tails.}
      \label{fig:rank-weight}
     
  \end{figure}

For OCV-aware timing analysis, delays at the \textit{tail sinks} are important for accurate skew prediction. We therefore propose a \emph{rank-weighted absolute error} loss that assigns higher importance to sinks at both extremes of the ground-truth delay distribution as shown in Fig.~\ref{fig:rank-weight}. 
  For a graph $g$ with sink count $|\mathcal{S}_g|$ and ground-truth delays $\{d_i\}_{i=1}^{|\mathcal{S}_g|}$ and predicted delays
  $\{\hat{d}_i\}_{i=1}^{|\mathcal{S}_g|}$, the normalized rank of each sink is:
  \begin{equation}
      r_i = \frac{\mathrm{rank}(d_i)}{|\mathcal{S}_g| - 1}, \quad r_i \in [0,\, 1]
      \label{eq:rank}
  \end{equation}
  where $\mathrm{rank}(d_i) \in \{0, 1, \ldots, |\mathcal{S}_g|{-}1\}$ is the ordinal
  position of sink~$i$ when sinks are sorted by increasing ground-truth
  delay.  We then define a U-shaped weighting function:
  \begin{equation}
      w(r_i) = 1 + \alpha \cdot 4\!\left(r_i - 0.5\right)^2
      \label{eq:weight}
  \end{equation}
  where $\alpha$ is a hyperparameter that controls the tail emphasis.  This quadratic function is minimized at the median ($r_i = 0.5$, where $w = 1$) and maximized at both extremes ($r_i \in \{0, 1\}$, where $w = 1 + \alpha$).  In our experiments, choosing $\alpha = 5$ gave us the best results, by which sinks at the tails receive $6\times$ the weight of median sinks, creating a smooth gradient of emphasis from center to extremes. The delay loss is then the weighted Mean Absolute Error (MAE):
  \begin{equation}
      \mathcal{L}_{\text{delay}} = \frac{1}{|\mathcal{S}_g|}\sum_{i=1}^{|\mathcal{S}_g|}
          w(r_i)\,\bigl|\hat{d}_i - d_i\bigr|
      \label{eq:delay_loss}
  \end{equation}
  We use $\ell_1$ (absolute error) as the error metric because it is more robust to occasional large outliers in SPICE-simulated labels, preventing a single anomalous sink from dominating the gradient.

  All other losses remain the same as in Section~\ref{sect:model-training}. We do not apply rank-based weighted loss for slew since it does not depend on the max and min values at sinks like clock skew.

\begin{figure}
    \centering
    \includegraphics[width=1.0\linewidth]{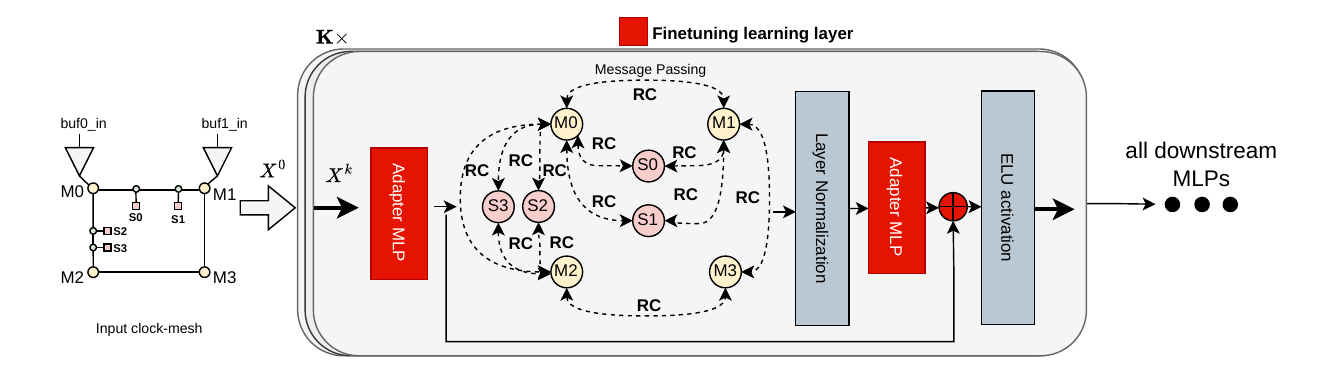}
    \caption{Finetuning of the base model in the Active Learning loop. The GNN message passing layers are frozen and two additional Pre-Message and Post-Message MLPs are added to finetune to the data. The final embeddings are passed to the downstream MLPs in a similar manner as in Fig.~\ref{fig:model}.}
    \label{fig:finetuning-model}
\end{figure}
\subsection{Active Learning for Tail-End Skew Refinement}
As discussed in Section~\ref{sect:synth-data-gen}, we have $N=50$ synthetic designs each having $P=20$ buffering densities. So, running thousands of MC simulations for each of the $N\times P$ designs to include $3\sigma$ tail-end skews would require an exorbitant amount of time for SPICE label generation in addition to training. Therefore, we propose a surrogate-guided active learning approach where we utilize our baseline model trained in Section~\ref{sect:train-methodology} to search in a loop (Algorithm~\ref{alg:al}) that iteratively identifies, labels, and learns from the Top-$K$ variation patterns that produce maximum skews in a given mesh. Since the model already predicts the average case and we want to improve its performance on tail-end predictions, we use a parameter efficient finetuning approach  where we keep the backbone GNN encoder frozen and add additional adapter layers similar to AdapterGNN~\cite{adapter-gnn} as shown in Fig.~\ref{fig:finetuning-model}. This framework allows the model to fit to the new data without forgetting the already learned average case representations. 
\begin{algorithm}[t]
  \caption{Active Learning for Worst-Case Skew Refinement}
  \label{alg:al}
  \begin{algorithmic}[1]
  \Require Clock mesh graphs $\{G_1,\dots,G_L\}$ stratified over buffering density;
           pretrained surrogate $f_\theta$ (frozen backbone, trainable adapter);
           sample budget $M$; inference batch-size $B$, label budget $K$; rounds $R$;
           sigma multiplier $n_\sigma$
  \Ensure  Refined surrogate $f_\theta$

  \State $\mathcal{D}_{\text{aug}} \gets \emptyset$
  \For{round $r = 1, \dots, R$}
            \For{each graph $G_i$}
          \State $\mathbf{Z}_i \gets \Call{SampleVariation}{G_i,\; M,\; n_\sigma}$
      % \Comment{$M$ i.i.d.\ corners, Table~\ref{table:variation}}
  \State $\hat{\mathbf{q}} \gets f_\theta\bigl(G_i,\; \mathbf{Z}_i, \; B\bigr)$
      \Comment{$M/B$ batched forward pass; $D$ delay quantiles per corner}
  \State $\hat{\mathbf{s}} \gets \hat{\mathbf{q}}_{D} - \hat{\mathbf{q}}_{1}$
      \Comment{predicted skew per corner}
          \State $\mathcal{Z}^{*}_i \gets \textsc{Top-}K\bigl(\mathbf{Z}_i,\; \hat{\mathbf{s}}\bigr)$
              \Comment{corners the surrogate deems worst}
          \ForAll{$\mathbf{z}^{*} \in \mathcal{Z}^{*}_i$} \textbf{in parallel}
              \State $y^* \gets \Call{RunSpice}{G_i,\; \mathbf{z}^{*}}$
              \State $\mathcal{D}_{\text{aug}} \gets \mathcal{D}_{\text{aug}}
                      \cup \bigl\{\bigl(G_i, y^*\bigr)\bigr\}$
          \EndFor
      \EndFor

      \State $f_\theta \gets \Call{Train}{f_\theta,\; \mathcal{D}_{\text{aug}}}$
          \Comment{adapter only; backbone frozen, train with Tail-Aware Loss}
  \EndFor
  \State \Return $f_\theta$
  \end{algorithmic}
  \end{algorithm}

  The procedure shown in Algorithm~\ref{alg:al} takes as input the following: a pool of $L$ mesh designs
  stratified over buffering density from our synthetic dataset, a pretrained surrogate
  $f_\theta$ from Section~\ref{sect:train-methodology} with the global GNN encoder frozen and adapter layers left trainable, and three budgets: $M$ variation samples per design, Top-$K$ configurations labeled per design,
  and $R$ refinement rounds.
  
  Each round begins by acquiring candidate configurations. For every design, line~4 draws
  $M$ variation samples in a single step. Line~5 then evaluates the surrogate on the $M$ variation samples and reads their predicted skew from the quantile head as $\hat{s}=\hat{q}_D - \hat{q}_1$, which is the predicted spread between the earliest and latest sink. Because all $M$ samples share a single topology and differ only in their node and edge feature values, the edge index that defines the graph connectivity is constructed once per design and reused, and the samples are evaluated in batches of $B$ rather than one at a time. Line~7 ranks
  the variations by predicted skew and retains the $K$ largest.
  
  The retained variation configurations are then labeled and the model is updated. Lines~8-11 invoke
  SPICE on each of the $K$ selected variation configurations, obtaining per-sink delays, slews and power, and append the resulting feature-label pairs to $\mathcal{D}_{\text{aug}}$. After the design loop closes (line~12), the round has
  contributed $L \times K$ labeled samples, and line~13 refits the adapter on the
  entire accumulated set rather than on the current round's additions alone, so
  that earlier acquisitions are not forgotten and the next round's ranking is
  produced by a strictly better informed model. The \Call{Train}{} then uses the Tail-Aware loss function we discussed in Section~\ref{sect:tail-loss} to train the delay head. All other heads are trained in a similar manner as described in Section~\ref{sect:model-training}.

  We set $R=4$, $L=25$, $M=5000$, $B=32$, and $K=5$. Using the fast surrogate model allows us to apply $M=5000$ variations for faster batched evaluation and pick the Top-$K$ skews. Although noisy, the surrogate refines in each round and we use the latest trained model before the start of the next round. The AL training loop performs additional $25\cdot 4\cdot5=500$ SPICE simulations for label generation. The whole AL methodology takes $10.03$~h total where $9.12$~h are spent on labeling the Top-$K$ configurations and $0.91$~h for training.  

\begin{table}[t]
    \centering
    \caption{Statistics of the test dataset that we evaluate our model on}
    \begin{tabular}{c|rrrr}
    \toprule
    \multirow{2}{*}{\textbf{Synthetic Benchmarks}} & \multicolumn{3}{|c}{\textbf{Stats}}\\
    & \textit{Area ($\mu m^2$)} & \textit{Sinks} & \textit{Grid Size} & \textit{Blockages}\\
    \midrule
    synth\_00 &  6294.0 & 5024 & 19x19 & 15 \\
    synth\_01 & 6197.0 & 1082 & 10x10  & 6 \\
    synth\_02 & 2846.9 & 1661 & 12x12 & 5 \\
    synth\_03 & 3795.6 & 1858 & 12x12 & 0 \\
    synth\_04 & 31063.4 & 13660 & 33x33 & 3 \\
    synth\_05 & 12770.6 & 7522 & 26x26 & 17 \\
    \midrule
    \textbf{Real Benchmarks} & \textit{Area ($\mu m^2$)} & \textit{Sinks} & \textit{Grid Size} & \textit{Blockages} \\
    \midrule
    riscv32i & 5569.9 & 1056 & 10x10 & 5 \\
    mock-alu & 2892.71 & 1413 & 12x12 & 0 \\
    ibex & 5658.32 & 1932 & 12x12 & 0 \\
    jpeg & 20585.92 & 4383 & 20x20 & 0 \\
    cva6 & 23960.2 & 4862 & 20x20 & 12\\
    coralnpu & 112525.3 & 8269 & 26x26 & 4 \\
    floonoc & 60421.6 & 13748 & 34x34 & 0\\
    
    \bottomrule
    \end{tabular}
    \vspace{-0.4cm}
    \label{tab:test-data-stats}
\end{table}
\section{Experimental Setup}\label{sect:experiment-setup}
We implemented our surrogate model in Pytorch~\cite{pytorch} with the Pytorch Geometric library~\cite{pytorch_geometric}. The Synopsys HSPICE and FM method as well as the benchmark generation were run on an AMD EPYC 7542 32-Core Processor (64 threads) machine with 512Gb of memory. Training and testing of our model was done on an NVIDIA GeForce RTX 4090 24GiB card. In order to ensure no data leakage from training dataset to test set we regenerate the synthetic benchmarks in Table~\ref{tab:test-data-stats} using the process outlined in Algorithm~\ref{alg:synth-gen}. The real benchmarks are taken from OpenROAD~\cite{openroad} and HighTide benchmarks~\cite{hightide-bench}. Clock meshes are implemented on the designs using ASAP7 7nm technology. The synthetic meshes are buffered randomly with uniform sampling from the PDK provided buffer library along with uniform buffering density sampling in range [0.1, 1.0]. Real design meshes are generated with bufferings from set-cover heuristics~\cite{pengli:combinatorial}. All the meshes are constructed with minimum total wire length~\cite{meshworks} and operate on a $1$~GHz frequency with $0.7$~V nominal supply voltage. For all experiments we apply OCVs respecting the realistic bounds in Table~\ref{table:variation}.
\subsection{Baseline Choices}\label{sect:baseline-choices}
As discussed in Section~\ref{sect: related-work-fm} our primary baseline to compare against is the FM method due to it being the only method that represents a clock mesh structure inherently and supports fast evaluations.

% \subsection{FM baseline}
The FM method is implemented following MeshWorks~\cite{meshworks} by forming a DC equivalent circuit
and constructing a graph Laplacian $\mathbf{G}$ where diagonal entries are sum of connected conductances and off-diagonal entries are negative of the branch conductance. The nodal analysis equation is then given by
\begin{math}
     \mathbf{G}v = I 
\end{math}
where $v$ is the resulting nodal voltages and $I$ is the vector of current sources. Since $\mathbf{G}$ is symmetric positive definite, we can solve this efficiently using Cholesky factorization. The resulting vector $v$ provides the first-order delay annotations, but does not provide slew.

A notable weakness of FM is that it assumes every mesh buffer switches
ideally at time zero, the nominal condition under which state-of-the-art
works employ it. Under OCV each buffer actually switches at a different time
and with a finite slew, and FM sees none of this. Since every sink is
driven jointly by many buffers through the low-resistance grid, the timing
that FM misses shifts all sinks of a given corner by nearly the same amount,
so its delay error is dominated by a single constant per corner rather than
by per-sink mistakes. To avoid understating the baseline, we also report
\textbf{FM+cal}, an idealized variant that is calibrated by removing the corner's mean sink
error computed from its own SPICE labels. This is the best any
constant correction can achieve. The large delay MAE reduction of FM+cal is
therefore expected by construction, and its residual isolates FM's true
per-sink accuracy. Skew, being the difference between the maximum and
minimum sink delays, cancels any constant shift and is identical under both
variants. We nevertheless report raw FM alongside FM+cal because the offset is not an artifact but a genuine consequence of FM's inability to represent differing input arrival times, and calibrating it away requires per-corner SPICE labels that no design-space exploration flow has. The two columns thus separate what FM gets wrong in principle from what remains after the most favorable correction.

% \subsection{\priorwork{}-new baseline}
We also compare against our prior work \priorwork{} which is the only GNN based method that represents the mesh structure inherently.

\section{Experimental Results}\label{sect:results}

\begin{figure}
    \centering
    \includegraphics[width=0.6\linewidth]{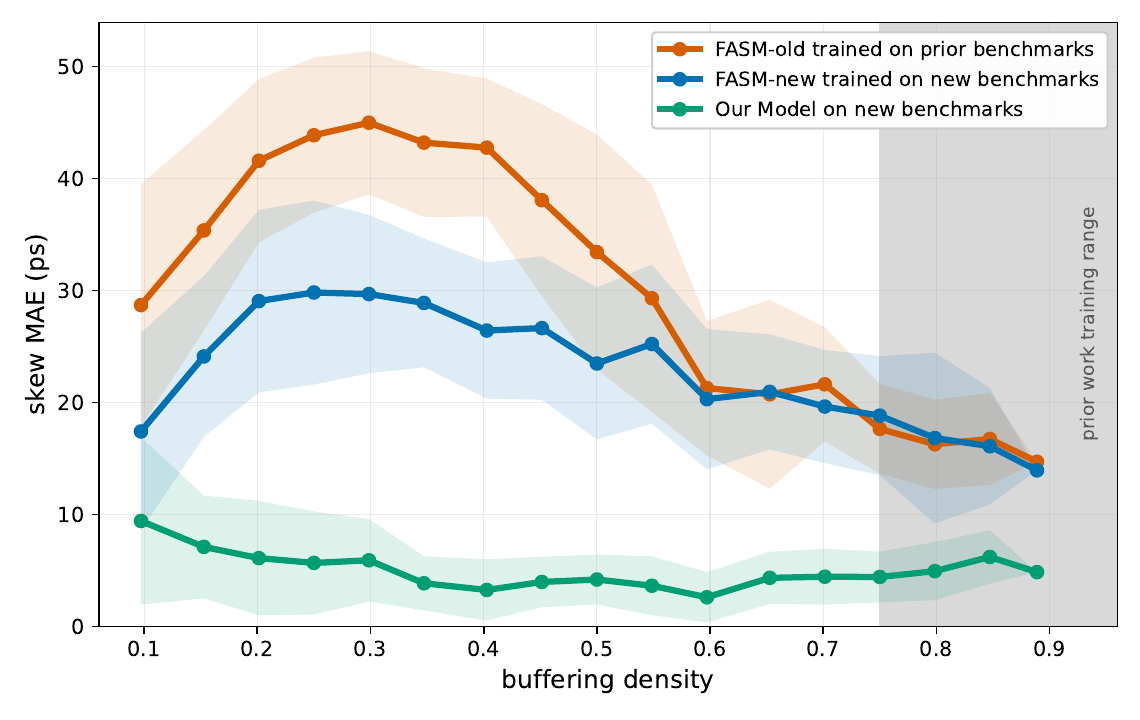}
    \caption{Comparison of different model inference capabilities on a buffering density search space. For each buffering density, we do $20$ random bufferings of \textit{ibex} and run SPICE to compare the model's inference with actual labels. Both the \priorwork{} models converge inside shaded prior work~\cite{khan:gatmesh} training range however, \priorwork-old is much more inaccurate for out-of-distribution bufferings as compared to \priorwork-new. Whereas, our proposed model in this work produces the least error in all the buffering density ranges.}
    \label{fig:synth-bench-comparison}
    \vspace{-0.3cm}
\end{figure}

\subsection{The need for new Synthetic Benchmark Generation}
The goal of a surrogate model for design-space exploration is to remain fairly constant in performance to different mesh buffering configurations in order to rank solutions. In this experiment we show why training on prior benchmarks is not a good idea anymore. In order to rank them, we train the model proposed in \priorwork{} on old synthetic benchmarks and call it \textbf{\priorwork-old} and train the same model on the new synthetic benchmarks and call it \textbf{\priorwork-new}. On a real design \textit{ibex}, we sweep the buffering density in range $[0.1, 0.9]$ and for each, apply $20$ random bufferings. The generated clock mesh variants are labeled by running SPICE and evaluated with \priorwork-old, \priorwork-new, and our proposed model in this work. The results are shown in Fig.~\ref{fig:synth-bench-comparison}. Both the \priorwork-old and \priorwork-new converge in prior work training range, whereas \priorwork-old is inaccurate at other densities which any search algorithm might explore to find better solutions. The \priorwork-new model performs considerably better however, our proposed model beats both the prior variants while maintaining fairly constant performance in all buffering densities. {\it Moving forward, we use \priorwork-new in all subsequent experiments for a fair comparison.}

\subsection{Graph Representation and Model Architecture Ablation}
\begin{figure}[t]
    \centering
    \includegraphics[width=1.0\linewidth]{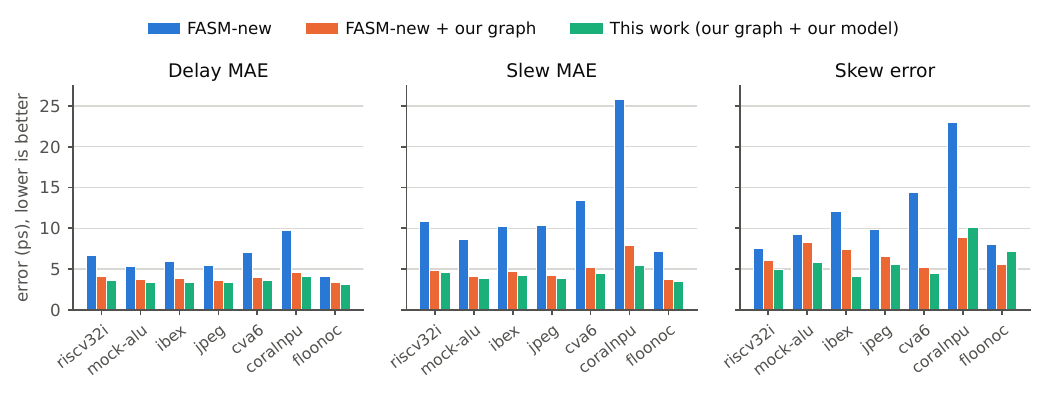}
    \caption{Ablation study showing individual improvements of the proposed graph representation and model architecture on the real designs. Each bar represents the mean error over $100$ Monte-Carlo corners per design.}
    \label{fig:ablation-model-graph}
\end{figure}

An ablation study is performed to highlight the individual enhancements of the proposed graph representation in Section~\ref{sect:graph-representation} and model architecture in Section~\ref{sect:model-architecture} against the old graph representation and model architecture in \priorwork. For this ablation we pick the real designs in Table~\ref{tab:test-data-stats} and apply $100$ random variations. The results are shown in Fig.~\ref{fig:ablation-model-graph}. The new graph representation itself provides significant improvements over the prior one in all test cases. The proposed model architecture further improves the performance on all but two test cases for skew error showing that combining both the graph representation and model architecture provides the best performance.

Furthermore, in Table~\ref{tab:ablation-graph-scalability} we show the scalability aspect of our graph representation. The total graph nodes and edges in \priorwork{} and the proposed graph representation are shown. On average we reduce the total nodes by $5.12\times$ and total edges by $3.87\times$. We show the peak memory allocation of the GPU when loading the graph and doing a single inference pass. Our proposed representation on average reduces the memory usage by $5.10\times$ thereby allowing more graphs to be loaded onto the GPU for batched inference. 
% ABLATION: graph representation vs model architecture
% G_o / G_n : old (wire_node) / new (wire_edge) graph representation
% M_o / M_n : old (delay+slew heads) / new (+delay-quantile, power heads) model

% GRAPH REPRESENTATION SCALABILITY ABLATION
\begin{table*}[t]
% \small
\centering
\caption{Graph representation scalability comparison with \priorwork{}. In the last row the node and edge entries are totals and their reduction is the combined ratio. Peak GPU memory is calculated over a single inference pass, measured with matched models so the difference is only due to the graph representation.}
\label{tab:ablation-graph-scalability}
\setlength{\tabcolsep}{4pt}
\begin{tabular}{lrrrrrrrrr}
\toprule
& \multicolumn{3}{c}{\textbf{Graph Nodes}} & \multicolumn{3}{c}{\textbf{Graph Edges}} & \multicolumn{3}{c}{\textbf{Peak GPU mem. (MB)}} \\
\cmidrule(lr){2-4}\cmidrule(lr){5-7}\cmidrule(lr){8-10}
\textbf{Design} & \priorwork{} & This Work & Red.\ & \priorwork & This Work & Red. & \priorwork{} & This Work & Red. \\
\midrule
riscv32i  &   5,724 & 1,128 & $5.07\times$ &   18,384 & 4,720 & $3.89\times$ & 280 & 55 & $5.09\times$ \\
mock-alu  &   8,001 &  1,557 & $5.14\times$ &   25,776 & 6,708 & $3.84\times$ & 395 & 77 & $5.13\times$ \\
ibex      &  10,596 & 2,076 & $5.10\times$ &   34,080 & 8,784 & $3.88\times$ & 521 & 102 & $5.11\times$\\
jpeg      &  24,595 & 4,783 & $5.14\times$ &   79,248 & 20,572 & $3.85\times$ & 1,206 & 237 & $5.09\times$ \\
cva6      &  26,237 &  5,157 & $5.09\times$ &   84,320 & 21,624 & $3.90\times$ & 1,284 & 251 & $5.11\times$ \\
coralnpu  &  44,577 &  8,753 & $5.09\times$ &  143,296 & 36,740 & $3.90\times$ & 2,184 & 428 & $5.10\times$ \\
floonoc   &  76,628 & 14,904 & $5.14\times$ &  246,896 & 63,968 & $3.86\times$ & 3,758 & 738 & $5.09\times$ \\
\midrule
\textbf{total / mean} & 196,358 & 38,358 & $\mathbf{5.12\times}$ & 632,000 & 163,116 & $\mathbf{3.87\times}$ & 9,628 & 1,888 & $\mathbf{5.10\times}$ \\
\bottomrule
\end{tabular}
\end{table*}

% OCV RESULTS ON SYNTHETIC TEST DATA TABLE
  \begin{table}[t]
  % \centering
  \caption{Clock mesh timing metric prediction with this work,
  \priorwork-new~\cite{khan:gatmesh}, first-moment (FM)~\cite{cmesh-int,
  meshworks, clockmesh:dp}, and its idealized calibrated variant FM+cal
  (Section~\ref{sect:baseline-choices}) on the synthetic test set with one MC
  sample and HSPICE labels as golden reference. Skew is a sink-pair difference
  and is identical under FM and FM+cal. Also, FM provides no slew estimate so omitted from table.}
  \begin{center}
  \small
  \setlength\tabcolsep{2pt}
  \vspace{-0.3cm}
  \begin{tabular}{c|cccc|cc|cccc}
  \toprule
  \multirow{2}{*}{\textbf{Designs}} &\multicolumn{4}{|c|}{\textbf{Delay MAE (ps)}}&\multicolumn{2}{|c|}{\textbf{Slew MAE (ps)}}&\multicolumn{4}{|c}{\textbf{Skew error (ps)}}\\
  % \cline{2-13}
  & This Work & \priorwork-new & FM & FM+cal & This Work & \priorwork-new & This Work & \priorwork-new & FM & FM+cal \\
  \midrule
  synth\_00 & \textbf{5.11} & 6.48  & 74.07 & 9.52  & \textbf{5.40}  & 10.17 & \textbf{0.91}  & 20.51 & 21.32 & 21.32 \\
  synth\_01 & \textbf{6.59} & 14.76 & 63.96 & 9.46  & \textbf{10.93} & 25.12 & \textbf{7.41}  & 8.58  & 35.86 & 35.86 \\
  synth\_02 & \textbf{2.62} & 4.13  & 75.13 & 4.01  & \textbf{4.46}  & 12.74 & \textbf{7.39}  & 9.13  & 16.11 & 16.11 \\
  synth\_03 & \textbf{3.91} & 4.39  & 81.71 & 5.62  & \textbf{4.23}  & 9.75  & \textbf{11.11} & 14.56 & 35.28 & 35.28 \\
  synth\_04 & \textbf{3.26} & 5.12  & 75.79 & 5.98  & \textbf{4.20}  & 8.18  & \textbf{2.32}  & 9.46  & 41.25 & 41.25 \\
  synth\_05 & \textbf{3.73} & 6.41  & 76.16 & 11.00 & \textbf{7.65}  & 9.84  & \textbf{10.12} & 13.93 & 21.64 & 21.64 \\
  \midrule
  \textbf{Average Error (ps)} & \textbf{4.20} & 6.88 & 74.47 & 7.60 & \textbf{6.14} & 12.63 & \textbf{6.54} & 12.69 & 28.57 & 28.57 \\
  \bottomrule
  \end{tabular} \label{table:synth-test-results}
  \end{center}
  \vspace{-0.3cm}
  \end{table}

% POWER RESULTS TABLE
\begin{table}[t] 
\centering
\caption{Clock mesh power comparison of this work, Dynamic Power ($CV^2f$) and actual power from running HSPICE on the synthetic benchmarks.}
\begin{center}
\setlength\tabcolsep{4pt}
% \vspace{-0.3cm}
\begin{tabular}{c|ccc|cc} 
\toprule
\multirow{2}{*}{\textbf{Synthetic designs}} &\multicolumn{3}{|c|}{\textbf{Power (mW)}} & \multicolumn{2}{|c}{\textbf{Error (\%)}}\\
% \cline{2-10}
& This Work & SPICE & Dynamic Power & This Work & Dynamic Power  \\
\midrule
synth\_00 & 8.73 & 8.35 & 5.18 & \textbf{4.5\%} & 37.9\% \\
synth\_01 & 2.31 & 2.27 & 1.47 & \textbf{1.9\%} & 35.3\% \\
synth\_02  & 4.43 & 4.53 & 2.65 & \textbf{2.2\%} & 41.5\% \\
synth\_03 & 4.66 & 4.82 & 2.66 & \textbf{3.3\%} & 44.6\% \\
synth\_04 &  37.67 & 36.98 & 20.02 & \textbf{1.8\%} & 45.9\% \\
synth\_05 & 13.87 & 13.25 & 7.84 & \textbf{4.7\%} & 40.8\% \\

\midrule
\textbf{Average Error (\%)} & & & & \textbf{3.06\%} & 41.0\% \\
\bottomrule
\end{tabular} \label{table:synth-power-results}
\end{center}
\vspace{-0.3cm}
\end{table}

\subsection{Model Performance on Synthetic Benchmarks}
For synthetic designs we apply a single random variation sample per design at a fixed $0.75$ buffering density. The results are shown in Table~\ref{table:synth-test-results}. Our model outperforms all baselines on
every metric. FM has significant accuracy loss where its raw delay MAE of $74.5$\,ps is dominated by the constant per-corner offset
discussed in Section~\ref{sect:baseline-choices}, due to its inability to
represent the input arrivals and slews that OCV on the top-level tree imposes
on the mesh buffers. The FM+cal column isolates what remains after that
offset is removed. Even with the idealized per-corner calibration, FM+cal
trails our model on delay ($7.60$ vs.\ $4.20$\,ps).
Skew is a relative difference between the maximum and minimum delay at the sinks so calibration does not affect it. FM
under-predicts the skew of every design, with a mean error of $28.6$\,ps
against $6.5$\,ps for our model. We further examine this at scale, when evaluating over
thousands of MC corners on real designs in Section~\ref{sect:results-real-bench}.

The power analysis results are shown in Table~\ref{table:synth-power-results}. The dynamic power estimate has on average $41.0\%$ error whereas, our model captures the actual SPICE computed power with significant accuracy with average error of $3.06\%$.  This shows that the global features we used along with our architecture helps the model learn the power consumption patterns in a clock mesh.

\begin{table}[t]
\small
\centering
% \caption{OCV model accuracy across training components. We report Mean Absolute Error (MAE) for delay and slew, average skew error, and tail skew error by measuring the $\text{99.9}^{\text{th}}$ percentile (P99.9) computed across \textbf{5000} MC runs per design.}
\caption{OCV model accuracy on the real designs, evaluated on \textbf{5000}
  SPICE-simulated MC corners per design. We report per-sink
  Mean Absolute Error (MAE) for delay and slew, and two distribution-level skew
  metrics: the average skew error, $|\text{mean(skew}_{\text{pred}})-~\!\text{mean(skew}_{\text{SPICE}})|$, 
  and the tail skew error, $|\text{P99.9}_{\text{pred}} - \text{P99.9}_{\text{SPICE}}|$, computed over
  each design's skew distribution. Our model is shown in three variants: trained
  without active learning (w/o AL), with AL under the default MSE loss (AL w/o Tail-Loss), and
  with AL under the proposed Tail-Loss (AL w/ Tail-Loss). FM+cal is the idealized
  per-corner calibrated FM of Section~\ref{sect:baseline-choices}. Skew is a
  sink-pair difference and is therefore identical under FM and FM+cal, and FM
  provides no slew estimate (--).}
\label{tab:ocv_ablation}
\renewcommand{\arraystretch}{1.15}
\setlength{\tabcolsep}{4pt}
\begin{tabular}{l|l|c c c | c | c | c}
\toprule
\multirow{2}{*}{\textbf{Designs}} & \multirow{2}{*}{\textbf{Metric (ps)}} & \multicolumn{3}{c|}{\textbf{This Work}}& \multirow{2}{*}{\textbf{\priorwork-new}} & \multirow{2}{*}{\textbf{FM}} & \multirow{2}{*}{\textbf{FM+cal}} \\
& & w/o AL & AL w/o Tail-Loss & AL w/ Tail-Loss & & &\\
 \midrule
\multirow{4}{*}{\textbf{riscv32i}}
  & Delay MAE          & 3.58 & \textbf{3.40} & 3.69 & 6.40 & 74.44 & 5.30\\
  & Slew MAE           & 4.53 & 4.46 & \textbf{4.27} & 10.98 & -- & -- \\
  & Skew Error (avg.)  & 2.07 & \textbf{0.32} & 1.81 & 5.57 & 25.53 & 25.53 \\
  % & Skew P95 Error     & 3.14 & \textbf{0.48} & 3.29 & 1.85 & 39.78  \\
  & Skew P99.9 Error     & 10.94 & 3.51 & \textbf{0.82} & 3.95 & 54.78 & 54.78   \\
  \midrule
\multirow{4}{*}{\textbf{mock-alu}}
  & Delay MAE          & 3.41 & \textbf{3.36} & 3.59 & 5.18 & 75.63 & 3.63  \\
  & Slew MAE           & 4.02 & 4.08 & \textbf{3.92} & 8.48 & -- & --  \\
  & Skew Error (avg.)  & 4.34 & 4.15 & \textbf{3.53} & 8.98 & 17.90 & 17.90 \\
  % & Skew P95 Error     & 2.98 & 2.74 & \textbf{2.34} & 7.83 & 26.04  \\
  & Skew P99.9 Error     & 6.20 & 0.24 & \textbf{0.08} & 6.70 & 37.46 & 37.46  \\
  \midrule
\multirow{4}{*}{\textbf{ibex}}
  & Delay MAE          & 3.31 &     \textbf{3.17} &  3.53     &     5.99 &      74.71 & 5.07 \\
  & Slew MAE           & 4.18 &        4.01  &  \textbf{3.91} & 10.24  & -- & -- \\
  & Skew Error (avg.)  & \textbf{0.56} & 1.39 & 2.85 & 10.93 & 26.04 & 26.04 \\
  % & Skew P95 Error     & 4.85 & 1.44 & 2.02  & 7.42  & 37.81 \\
  & Skew P99.9 Error     & 10.22 & 5.57 & \textbf{1.18} & 5.56  &  49.83 & 49.83 \\

\midrule
\multirow{4}{*}{\textbf{jpeg}}
  & Delay MAE          & 3.07 & \textbf{2.98} & 3.00 & 5.38 & 74.99 & 5.11\\
  & Slew MAE           & 3.85 & 4.05 & \textbf{3.83} & 10.09 & --  & -- \\
  & Skew Error (avg.)  & \textbf{3.96} & 5.49 & 5.32 & 9.30 & 27.07 & 27.07 \\
  % & Skew P95 Error     & \textbf{0.50} & 2.98 & 3.97 & 7.85 & 36.95  \\
  & Skew P99.9 Error     & 6.62 & \textbf{0.29} & 0.87 & 7.43 & 46.77 & 46.77   \\
\midrule
\multirow{4}{*}{\textbf{cva6}}
  & Delay MAE          & 3.23 & \textbf{3.19} & 3.47 & 7.05 & 73.90 & 6.20\\
  & Slew MAE           & 4.49 & 4.43 & \textbf{4.29} & 13.48 & -- & --  \\
  & Skew Error (avg.)  & 1.05 & 2.58 & \textbf{0.42} & 14.40 & 32.81 & 32.81 \\
  % & Skew P95 Error     & 5.99 & 6.50 & \textbf{3.12} & 11.31 & 44.86  \\
  & Skew P99.9 Error     & 9.88 & 8.85 & \textbf{7.79} & 9.50 & 57.23 & 57.23   \\

    \midrule
\multirow{4}{*}{\textbf{coralnpu}}
  & Delay MAE          & 3.53 & \textbf{3.47} & 3.51 & 9.85 & 71.29 & 7.54 \\
  & Slew MAE           & \textbf{5.26} & 5.67 & 5.30 & 25.99 & -- & --  \\
  & Skew Error (avg.)  & 9.79 & 13.43 & \textbf{9.69} & 23.53 & 34.91 & 34.91  \\
  % & Skew P95 Error     & 2.79 & 9.40 & 7.44 & 19.51 & 45.44  \\
  & Skew P99.9 Error     & 5.28 & 3.65 & \textbf{2.74} & 17.42 & 56.58 & 56.58   \\
  \midrule

\multirow{4}{*}{\textbf{floonoc}}
  & Delay MAE          & 2.92 & \textbf{2.79} & 3.07 & 4.10 & 75.61 & 4.45\\
  & Slew MAE           & 3.54 & 3.79 & \textbf{3.49} & 7.20 & -- & -- \\
  & Skew Error (avg.)  & \textbf{6.52} & 9.72 & 7.61 & 8.60 & 28.31 & 28.31 \\
  % & Skew P95 Error     & 6.36 & 1.21 & \textbf{2.13} & 5.18 & 36.29  \\
  & Skew P99.9 Error     & 6.48 & 3.42 & \textbf{1.16} & 6.50 & 45.29 & 45.29   \\

\midrule
\textbf{Average Delay MAE} & & 3.29 & \textbf{3.19} & 3.40 & 6.28 & 74.37 & 5.33\\
\midrule
\textbf{Average Slew MAE} & & 4.27 & 4.36 & \textbf{4.14} & 12.35 & -- & -- \\
\midrule
\textbf{Average Skew Error} & & \textbf{4.04} & 5.30 & 4.46 & 11.62 & 27.51 & 27.51 \\
\midrule
\textbf{Average Skew P99.9 Error} & & 7.95 & 3.65 & \textbf{2.09} & 8.15 & 49.70 & 49.70 \\
\bottomrule 
\end{tabular}
\vspace{-0.3cm}
\end{table}

\subsection{Model Performance on Real Benchmarks}\label{sect:results-real-bench}
For extensive performance analysis on real benchmarks, we evaluate every model on 5000 SPICE-simulated MC corners per design. To attribute our contributions individually, Table~\ref{tab:ocv_ablation} reports three variants of our model: trained on the base synthetic dataset without AL (w/o AL, Section~\ref{sect:train-methodology}), with AL under the default MSE loss (AL w/o Tail-Loss) and with AL under proposed Tail-Loss (AL w/ Tail-Loss). The w/o AL variant already matches the others on mean metrics but degrades at the tail. With only $M=5$ MC samples per training mesh, tail-producing variation patterns are rare in the base training data (Section~\ref{sect:synth-data-gen}), and its $\text{P}99.9$ skew error reflects that. AL improves $\text{P}99.9$ on $7/7$ designs, and Tail-Loss improves it further on $6/7$ at a slight cost in average delay MAE ($3.19$~ps to $3.40$~ps). The full model (AL w/ Tail-Loss) outperforms \priorwork-new on every design and every metric.

\begin{figure}[t]
    \centering
    % Row 1
    \begin{subfigure}{0.8\textwidth}
        \centering
        \includegraphics[width=\linewidth]{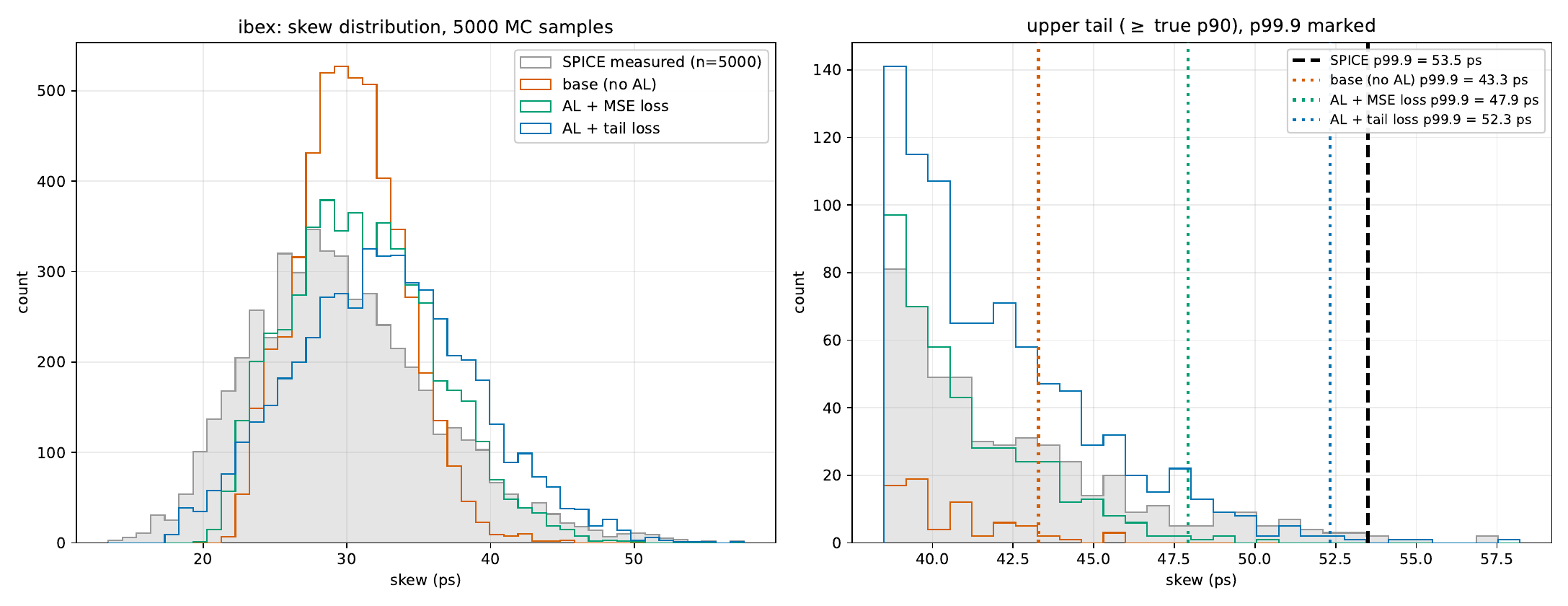}
        % \caption{First image}
        \label{fig:sub1}
    \end{subfigure}
    \vspace{-0.5cm} % vertical space between rows
    
    % Row 2
    \begin{subfigure}{0.8\textwidth}
        \centering
        \includegraphics[width=\linewidth]{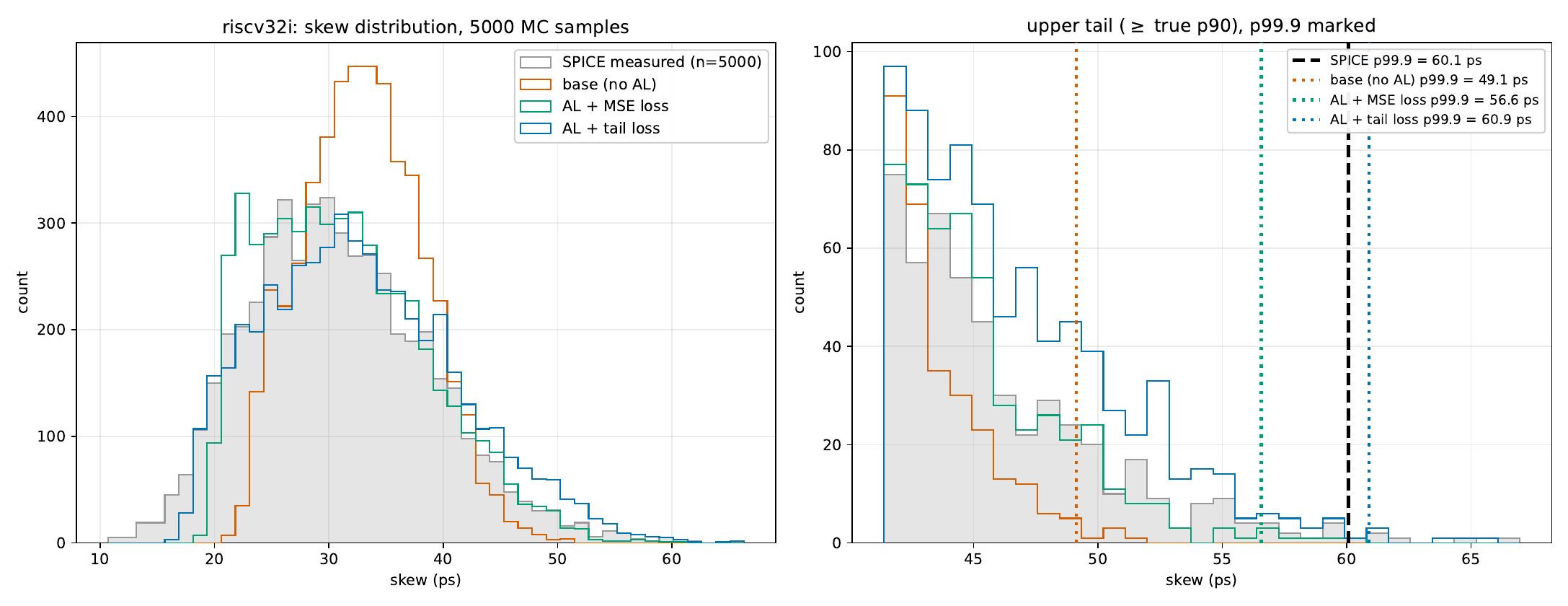}
        % \caption{Third image}
        \label{fig:sub3}
    \end{subfigure}
    
    \caption{Skew histogram plots of the two benchmarks (ibex, riscv32i). \textbf{Left:} plots shows the overall skew histogram produced by SPICE, and our three model variants. \textbf{Right:} plots shows the zoomed in tail-end of the skew histogram and how each model captures it.}
    \label{fig:skew-hist}
   \vspace{-0.5cm}
\end{figure}

The FM columns depict the analytical baseline's structural limits. The FM without calibration carries a large, nearly constant per-corner delay offset (Section~\ref{sect:baseline-choices}). Once the idealized
calibration removes this offset, FM+cal surpasses \priorwork-new on delay MAE
($5.33$~ps vs.\ $6.28$~ps on average), but it still remains $1.6\times$ above our full model
($5.33$~ps vs.\ $3.40$~ps). Skew is not affected by the delay offset since it is a local relative change between the maximum and minimum delays at sinks. FM under-predicts the skew across all seven designs, with an
average error of $27.51$~ps and a P$99.9$ error of $49.70$~ps. In
contrast, our full model reduces the average P$99.9$ skew error by $24\times$
relative to FM ($49.70$~ps vs.\ $2.09$~ps) and by $3.9\times$ relative to
\priorwork-new ($8.15$~ps vs.\ $2.09$~ps). Fig.~\ref{fig:skew-hist} visualizes this
behavior by plotting each model's skew histogram against SPICE, showing where
the mean and the tail of the distribution are captured or lost.

We also present our model's performance on power prediction in Table~\ref{table:power-results}. Our model generalizes to the real benchmarks as well and shows good accuracy in power prediction.

% HISTOGRAM PLOTS 

% POWER RESULTS TABLE
  \begin{table}[t]
  \centering
  \caption{Clock mesh power comparison of this work, the dynamic
  power estimate ($CV^2f$), and actual HSPICE power on real designs, averaged over 5000 MC
  corners per design. Error is the mean per-corner absolute percentage error
  against SPICE.}
  \begin{center}
  \setlength\tabcolsep{4pt}
  \begin{tabular}{c|ccc|cc}
  \toprule
  \multirow{2}{*}{\textbf{Real Designs}} &\multicolumn{3}{|c|}{\textbf{Power (mW)}} & \multicolumn{2}{|c}{\textbf{Error (\%)}}\\
  & This Work & SPICE & Dynamic Power & This Work & Dynamic Power \\
  \midrule
  riscv32i & 2.38  & 2.45  & 1.44  & \textbf{3.8\%} & 41.2\% \\
  mock-alu & 3.47  & 3.36  & 1.88  & \textbf{3.7\%} & 44.1\% \\
  ibex     & 4.69  & 4.66  & 2.69  & \textbf{2.0\%} & 42.3\% \\
  jpeg     & 10.36 & 10.00 & 5.89  & \textbf{3.8\%} & 41.1\% \\
  cva6     & 9.54  & 9.21  & 5.75  & \textbf{3.7\%} & 37.6\% \\
  coralnpu & 16.35 & 15.71 & 10.84 & \textbf{4.1\%} & 31.0\% \\
  floonoc  & 29.81 & 29.13 & 16.84 & \textbf{2.4\%} & 42.2\% \\
  \midrule
  \textbf{Average Error (\%)} & & & & \textbf{3.4\%} & 39.9\% \\
  \bottomrule
  \end{tabular} \label{table:power-results}
  \end{center}
  \end{table}

\begin{table*}[t]
  \setlength\tabcolsep{4pt}
  \caption{Runtime for one mesh evaluation and for a 5000-sample Monte-Carlo
  sweep. We denote the SPICE netlist to graph conversion time as~(\emph{graph conv.})\ and model inference time as~(\emph{infer.}). Under variation our model reuse a single converted
  graph and update only its features, so conversion is paid once. FM must be
  recomputed per corner.}
  \centering
  \begin{tabular}{l|rrrrrr|rr|rr}
  \toprule
  \multirow{3}{*}{\textbf{Design}}
    & \multicolumn{6}{c|}{\textbf{Runtime (s)}}
    & \multicolumn{4}{c}{\textbf{Speedup of This Work}} \\
  \cmidrule{2-7}\cmidrule{8-11}
    & \multirow{2}{*}{HSPICE} & \multirow{2}{*}{FM}
    & \multicolumn{2}{c}{This Work} & \multicolumn{2}{c|}{\priorwork}
    & \multicolumn{2}{c|}{over FM} & \multicolumn{2}{c}{over \priorwork} \\
  \cmidrule(lr){4-5}\cmidrule(lr){6-7}\cmidrule(lr){8-9}\cmidrule(lr){10-11}
    & & & graph conv. & infer. & graph conv. & infer.
    & 1 pass & 5000 MC & 1 pass & 5000 MC \\
  \midrule
  riscv32i & 1.86   & 0.15  & 0.1781 & 0.0082 & 0.1999 & 0.0095 & 0.8$\times$ &   18$\times$ & 1.12$\times$ & 1.16$\times$ \\
  mock-alu & 4.92   & 0.37  & 0.2564 & 0.0038 & 0.2903 & 0.0088 & 1.4$\times$ &   96$\times$ & 1.15$\times$ & 2.30$\times$ \\
  ibex     & 8.67   & 0.51  & 0.3550 & 0.0052 & 0.4032 & 0.0116 & 1.4$\times$ &   97$\times$ & 1.15$\times$ & 2.22$\times$ \\
  jpeg     & 28.03  & 3.56  & 0.8872 & 0.0069 & 1.3250 & 0.0269 & 4.0$\times$ &  503$\times$ & 1.51$\times$ & 3.84$\times$ \\
  cva6     & 30.32  & 2.80  & 1.3490 & 0.0120 & 1.4435 & 0.0275 & 2.1$\times$ &  228$\times$ & 1.08$\times$ & 2.26$\times$ \\
  coralnpu & 72.42  & 7.41  & 2.3416 & 0.0142 & 2.8020 & 0.0405 & 3.1$\times$ &  505$\times$ & 1.21$\times$ & 2.80$\times$ \\
  floonoc  & 166.53 & 37.20 & 4.4951 & 0.0209 & 5.3001 & 0.0792 & 8.2$\times$ & 1706$\times$ & 1.19$\times$ & 3.68$\times$ \\
  \bottomrule
  \end{tabular}
  \label{tab:runtime}
  \end{table*}

\subsection{Runtime Evaluation}

 The key reason to use FM method in state-of-the-art works is the speedup gain. However, the trade-off is the accuracy loss as we saw in Table~\ref{tab:ocv_ablation}. In order to measure our runtime gains with the FM method and with our prior work~\priorwork{}, we compare them over a single pass and over $5000$ MC sample evaluations as shown in Table~\ref{tab:runtime}. In one-pass speedup comparison we have better performance on all designs as compared to FM except the smallest design (riscv32i, $0.8\times$) where graph construction dominates. However, we are faster in all the designs from our prior work~\priorwork{} which is mainly due to our new graph representation of the clock mesh. The major advantage we get over FM is in multiple calls during OCV evaluations, where we remain in the graph domain and update the features of the graph to represent OCV thereby requiring only the inference for each evaluation. On the contrary, FM method needs to compute the delays again.

\section{Case Study: Variation-Aware Buffer Sizing}\label{sect:case-study}

In this section, we present a study on using our proposed surrogate model inside an optimization loop. Here all the presented components of the model are utilized such as skew, slew, and power prediction under OCV. {\it Using SPICE alone and not a surrogate model, this optimization would not be possible due to run-time.} For this specific study, we use a multi-objective optimization method based on genetic algorithm called NSGA-II~\cite{nsga-ii} where we optimize the skew, max slew and power under OCV. Our approach is not tied to NSGA-II since the surrogate serves only as a fast evaluator.

For this experiment, we select the \textit{ibex} design with operating frequency of $1$~GHz. We provide the NSGA-II algorithm with the clock mesh where each intersecting mesh point is the candidate for a buffer to be placed. In the case of \textit{ibex} we have a grid size of $12\times12$ (see Table~\ref{tab:test-data-stats}) meaning $12\cdot12=144$ potential buffering candidate locations. Each buffering candidate is represented as a vector $\vec{b}$ of different buffer sizes. We use the ASAP7 PDK provided legal buffer sizes of $\{0,2,3,4,5,6,8,10,12,16,24\}$ where size $0$ just means no buffer at the location. The candidates vectors are reflected in the graph where the buffer related features are modified accordingly and then $M$ random variations are applied. Each candidate and their variations are independent and are evaluated in batched forward pass through the model. The candidate is scored on three objectives: the $99^{th}$ percentile of predicted skew and max-slew, and the power consumption across the $M$ sampling variations. We used percentiles rather than the exact max value of the skew/slew across the $M$ variations due to randomness, where re-evaluating the same candidate with another $M$ random variations can produce a different max skew/slew value and also due to the surrogate model noise where it might over-predict or under-predict certain candidates at a single max value over all $M$ evaluations. Using percentiles gives the model more leverage to explore the candidates.

We initialize a population of $N_{pop}=100$ buffering candidates $\vec{b_1}, \vec{b_2}, ...,\vec{b_{100}}$ in two ways. First, we generate $50\%$ of the informed candidates encoding a physical prior: for a fixed total drive strength the aggregate output conductance of the buffer set is unchanged, but distributing it over more sites shortens the mesh wire resistance in series with each driver and therefore lowers the effective driver resistance seen at every node, improving slew at identical buffer area. These candidates are drawn from a buffering density between $[0.85, 1.00]$ and assign each active site a legal buffer size $s$ with probability proportional to $1/s^{2}$. This creates densely buffered meshes with many smaller buffers as opposed to sparse yet larger buffers yielding a mean buffer size of $3.5$. Second, for the remaining candidates we do exploration by uniformly sampling the buffering density in range $[0.10, 1.00]$ and uniformly sampling the buffer size from the library yielding a mean buffer size of $9.0$. With this approach of initialization, we provide the algorithm with a balance of prior knowledge and exploration.

The parents are selected by binary tournament on front rank and crowding distance, recombined by uniform crossover with $p=0.9$ and mutated per site with $p=0.03$ by a move that steps a buffer to the adjacent larger or smaller legal size, randomizes it, or removes it. Offspring are merged with their parents and the whole pool is re-sorted each generation. We run a total of $N_{gen}=20$ generations and use $M=5000$ random variation samples.

The algorithm returns $66$ Pareto front points that are evaluated with SPICE. The best candidates are chosen based on the minimum max-slew, minimum power, and one that balances all objectives.

\subsection{Buffering Optimization Results}

For comparison of our buffering solutions, we use the baseline buffered solution proposed by Venkataraman et al.~\cite{pengli:combinatorial} and with current state-of-the-art MeshWorks~\cite{meshworks}. The results are shown in Table~\ref{tab:buffering} and Table~\ref{tab:buffering_delta}.
The balanced design improves the $99^{\text{th}}$-percentile skew by $49.1\%$, power by
$25.7\%$ and buffer area by $37.3\%$ over MeshWorks for a $6.8\%$ slew increase.
Moving along the front trades skew against the remaining budgets: the lowest power design
gives up $4.6\%$ of skew improvement to gain a further $2.7\%$ of
power and $3.5\%$ of area, while the min-slew design reduces the slew penalty to
$4.4\%$ and still improves $99^{\text{th}}$-percentile skew by $40.5\%$ at MeshWorks'
exact buffer count. The heuristic proposed in baseline~\cite{pengli:combinatorial} attains
the best slew of any design considered ($105.3$~ps) but at the cost of $3.1\times$ the area and
$2.3\times$ the power of our balanced design.

\begin{table}[t]
  \centering
  \caption{Buffering optimization on \textit{ibex}, verified with 5000 HSPICE
  Monte Carlo corners per design at matched process seeds. Skew and maximum
  slew are reported as percentiles of the per-corner distribution, power is
  the mean.}
  \label{tab:buffering}
  \setlength{\tabcolsep}{4pt}
  \begin{tabular}{lrrrrrrrrr}
  \toprule
  & \multicolumn{3}{c}{skew (ps)} & \multicolumn{3}{c}{max.\ slew (ps)} & power & area & \# \\
  \cmidrule(lr){2-4}\cmidrule(lr){5-7}
  \textbf{ibex design} & $p_{50}$ & $p_{99}$ & $p_{99.9}$ & $p_{50}$ & $p_{99}$ & $p_{99.9}$ & (mW) & ($\mu$m$^2$) & bufs \\
  \midrule
  baseline~\cite{pengli:combinatorial}              & 29.1 & 48.2 & 53.5 & \textbf{94.5} & \textbf{105.3} & \textbf{108.6} & 4.71 & 54.37 & 128 \\
  MeshWorks~\cite{meshworks}       & 21.1 & 37.3 & 43.0 & 98.4 & 112.4 & 117.6 & 2.75 & 27.77 & 139 \\
  \midrule
  % Ours (random init.)              & 19.4 & 33.8 & 39.7 & 106.7 & 119.0 & 122.1 & 2.74 & 27.51 & 99 \\
  Ours (min.\ slew)        & 11.6 & 22.2 & 26.7 & 106.5 & 117.3 & 120.2 & 2.13 & 18.63 & 139 \\
  Ours (balanced)          & \textbf{10.2} & \textbf{19.0} & \textbf{22.8} & 108.7 & 120.0 & 124.0 & 2.04 & 17.41 & 138 \\
  Ours (min.\ power)       & 10.9 & 20.7 & 24.4 & 111.2 & 122.6 & 127.2 & \textbf{1.97} & \textbf{16.45} & 129 \\
  \bottomrule
  \end{tabular}
  \end{table}
\begin{table}[t]
  \centering
  \caption{Relative change of the optimized buffering solutions with respect to each
  baseline, on the metrics of Table~\ref{tab:buffering}. Skew and maximum slew are
  reported at the median ($p_{50}$) and the 99th percentile ($p_{99}$) over 5000
  matched Monte Carlo corners, power is the mean and area is total buffer cell area.
  \textbf{Positive values denote improvement.}}
  \label{tab:buffering_delta}
  \setlength{\tabcolsep}{8pt}
  \begin{tabular}{lrrrrrr}
  \toprule
  & \multicolumn{2}{c}{skew} & \multicolumn{2}{c}{max.\ slew} & & \\
  \cmidrule(lr){2-3}\cmidrule(lr){4-5}
  \textbf{ibex design} & $p_{50}$ & $p_{99}$ & $p_{50}$ & $p_{99}$ & power & area \\
  \midrule
  \multicolumn{7}{l}{\textit{vs.\ MeshWorks}~\cite{meshworks}} \\
  % Ours (random init.) &  $+8.1\%$ &  $+9.4\%$ &  $-8.4\%$ & $-5.9\%$ &  $+0.3\%$ &  $+0.9\%$ \\
  Ours (min.\ slew)  & $+45.0\%$ & $+40.5\%$ &  $-8.2\%$ & $-4.4\%$ & $+22.3\%$ & $+32.9\%$ \\
  Ours (balanced)    & $+51.7\%$ & $+49.1\%$ & $-10.5\%$ & $-6.8\%$ & $+25.7\%$ & $+37.3\%$ \\
  Ours (min.\ power) & $+48.3\%$ & $+44.5\%$ & $-13.0\%$ & $-9.1\%$ & $+28.4\%$ & $+40.8\%$ \\
  \midrule
  \multicolumn{7}{l}{\textit{vs.\ baseline}~\cite{pengli:combinatorial}} \\
  % Ours (random init.) & $+33.3\%$ & $+29.9\%$ & $-12.9\%$ & $-13.0\%$ & $+41.9\%$ & $+49.4\%$ \\
  Ours (min.\ slew)  & $+60.1\%$ & $+53.9\%$ & $-12.7\%$ & $-11.4\%$ & $+54.7\%$ & $+65.7\%$ \\
  Ours (balanced)    & $+64.9\%$ & $+60.6\%$ & $-15.0\%$ & $-14.0\%$ & $+56.7\%$ & $+68.0\%$ \\
  Ours (min.\ power) & $+62.5\%$ & $+57.1\%$ & $-17.6\%$ & $-16.4\%$ & $+58.3\%$ & $+69.8\%$ \\
  \bottomrule
  \end{tabular}
  \end{table}
  
% runtime results 
\subsection{Buffering Optimization Runtime}
Producing the Pareto front requires $N_{gen}\cdot N_{pop}\cdot M=20\cdot100\cdot5000$ resulting in $10^7$ evaluations, which completes in $5.5$~h on two NVIDIA RTX~4090s. SPICE is then run once per front point to validate the search, taking a further $9.5$~min for a $66$ point Pareto front. Selection of the candidates are done after running SPICE on the Pareto front points. Once the final candidates are selected, we run $5000$ MC simulations on each to compare their robustness to OCV against the other baselines.

Table~\ref{tab:runtime} reports single sequential runtimes, but both FM and HSPICE can be parallelized, so we compare against their best case on our machine. HSPICE corners are independent and scale across processes. On our
$64$-thread host, however, throughput saturates at $8$ concurrent HSPICE
instances because the runs are memory-bandwidth bound, and using more workers gains
nothing, giving an effective $1.1$\,s per ibex corner. Even at that rate the
$10^{7}$ evaluations would take roughly four months, about $550\times$ longer
than the surrogate. This showcases why HSPICE is infeasible in search space exploration. However, it remains as the only exact evaluator, which is why we reserve it for validating the front rather than searching for it.

FM is a more interesting case because its solves are also independent, and at $0.51$~s per ibex corner on
$32$ cores the $10^{7}$ evaluations would finish in under two days if it scales perfectly. FM is
therefore feasible but still falls short of the surrogate. However, it is not excluded on the runtime but rather due to its limitation and inaccuracies in computing metrics. It provides no slew for the feasibility constraint, the power
estimate is inaccurate ($42.3\%$ error for ibex), and its computed skew worsens in the tail-ends under OCV ($49.83$~ps error for ibex).

Our surrogate does carry a one-time cost, namely synthetic data generation and training, but
this cost is amortized: the model is inductive and is trained once for the
technology, not per design or per search. Among the three evaluators, only
the surrogate is simultaneously fast enough for the search and accurate
enough for its objectives.

\section{Conclusion}\label{sect:conclusion}
This work proposed the first full surrogate model of delay, slew, skew, power and variation for clock meshes and showed significant improvement in the timing metrics over its predecessor \priorwork{}. By modeling the clock mesh with an improved graph representation along with the new derived edge and global features, and a new model architecture the surrogate model effectively captures the complex interactions of the multiple drivers, reconverging paths, and OCV. Furthermore, with the new AL strategy along with our proposed Tail-Aware loss function we significantly improve capturing the skew histogram under OCV. Results show that the proposed model predicts all the metrics with marginal errors as compared to SPICE and significantly improves over the FM method used in current state-of-the-art works. We also motivate the use-case of a surrogate model to provide faster and accurate metrics at design-time for informed early decisions during clock mesh optimization and faster design-space exploration.

\bibliographystyle{ACM-Reference-Format}
\bibliography{main}

\end{document}